\documentclass[10pt,aps,pra,twoside,twocolumn]{revtex4-2}

\usepackage[english]{babel}
\usepackage{bm}
\usepackage{float}
\usepackage{dsfont}
\usepackage{braket}
\usepackage{amsmath}
\usepackage{amssymb}
\usepackage{fancyhdr}
\usepackage{graphicx}
\usepackage{mathtools}
\usepackage{changepage}
\usepackage[colorlinks=true, allcolors=black]{hyperref}

\usepackage{silence}
\newcommand{\avg}[1]{\langle #1 \rangle}
\newcommand{\cg}[6]{\langle #1\, #2,\, #3\, #4 | #5\, #6 \rangle}
\newcommand{\cgr}[6]{\langle #1\, #2,\, #3\, #4 \| #5\, #6 \rangle}
\newcommand{\Om}{\Omega}
\newcommand{\Oms}{\dim}
\newcommand{\hw}{\hbar\Omega}
\newcommand{\Nmax}{N_{\rm max}}
\newcommand{\Hb}{\mathcal{H}}
\newcommand{\Gb}{\mathcal{G}}

\newcommand{\ama}{a m_a}
\newcommand{\bmb}{b m_b}

\newcommand{\SU}[1]{\ensuremath{\mathrm{SU}( #1 )}}
\newcommand{\Un}[1]{\ensuremath{\mathrm{U}( #1 )}}
\newcommand{\SO}[1]{\ensuremath{\mathrm{SO}( #1 )}}

\newcommand{\SpR}[1]{\ensuremath{\mathrm{Sp}( #1,\mathbb{R} )}}
\newcommand{\one}{\mathds{1}}
 
\begin{document}

\title{Entanglement, correlations, and resource costs\\for symmetry-adapted quantum simulations of light through medium-mass nuclei}
\author{Darin C. Mumma$^{1}$}
\author{Eunjin In$^{1}$}
\author{Noah Thompson$^{1}$}
\author{Kristina D. Launey$^{1}$}
\affiliation{$^{1}$Department of Physics and Astronomy, Louisiana State University, Baton Rouge, LA 70803, USA}
\date{November 10, 2026}

\begin{abstract}

Entanglement governs both the classical cost of simulating correlated many-body systems and the quantum resources needed to treat them. In this work, we use nuclear symmetries to investigate the entanglement and correlations in nuclei using \textit{ab initio} symmetry-adapted no-core shell-model nuclear wavefunctions, illustrated here for ${}^{4,6,8}\mathrm{He}$, $^8$Be, ${}^{12}\mathrm{C}$, ${}^{20}\mathrm{Ne}$, and $^{48}$Ca. In this many-body approach, deformation and collectivity naturally emerge from the underlying chiral potential, and we show that the Sp(3,$\mathbb{R}$) basis concentrates the ground state in far fewer basis states than the SU(3) basis. We compute single-mode entropies and two-mode mutual information, and examine the organization of like-particle two-nucleon correlations into symmetry channels. Single-mode entropies are largest in the valence shells, while two-mode mutual information reveals substantial correlations both within and between shells. In the harmonic-oscillator basis, negativity identifies entangled pairs of modes. This entanglement vanishes in the natural-orbital basis and is what a circuit on an oscillator-basis register must generate. We also study the basis-independent linear entropy, which brackets the natural-orbital entropy without diagonalizing the one-body density, while informing state preparation. In all cases considered, the dominant pair correlations are captured by fewer SU(3) channels than angular-momentum channels, which reduces the number of variational parameters. Since the number of operator components within the channels is comparable in the two schemes for the heavier nuclei, this gives the SU(3)-coupling scheme a clear advantage. Fiedler ordering of the correlation graph with local refinement reduces the register ordering cost by a factor of 2 to 3 relative to the original shell ordering, and by 10 when two more shells are included in the register. Together, these measures link nuclear correlations to the selection of active modes and candidate excitation operators and to the design of quantum register layouts, offering guidance for the development of hybrid quantum-classical calculations of nuclei.

\end{abstract}
\maketitle
\tableofcontents

\section{Introduction}

Entanglement has emerged both as a lens on nuclear structure and as a measure of how hard a nucleus is to model, whether classically or on quantum hardware. \textit{Ab initio} calculations of $^4$He and $^6$He showed that mode entanglement depends strongly on the single-particle basis~\cite{Robin2021}. In valence-space shell-model calculations, single-orbital entropies track the number of valence nucleons and the shell structure, two-orbital mutual informations reveal like-particle pairing and deformation, and proton--neutron entanglement is consistently weak~\cite{PerezObiol2023,Johnson2023}. Analytic studies of two-nucleon and seniority-model states account for these patterns and show how isospin symmetry and the choice of $LS$- or $jj$-coupled bases shape the measures~\cite{Kruppa2021,Kruppa2022,Kovacs2025}. \textit{Ab initio} coupled-cluster, in-medium similarity renormalization group, and no-core shell-model calculations have since extended these analyses~\cite{Gu2023,Tichai2023,Sarma2026,Shinde2026IoI}. Entanglement entropy has also been linked to short-range correlations~\cite{Bulgac2023,Pazy2023} and proposed as a signature of shell evolution and shape change~\cite{Xu2026,Wang2026}. Beyond bipartite measures, $n$-tangles and non-stabilizerness (magic) quantify multinucleon complexity, revealing substantial proton--neutron-dominated multinucleon entanglement in the $sd$ shell and shell-closure signatures in the $pf$ shell~\cite{Hengstenberg2023,Brokemeier2025,Robin2025,ShindeSrivastava2026}. Low entanglement has, in turn, become a computational resource: the weakness of proton--neutron entanglement underlies a shell-model truncation that reaches model spaces beyond standard codes~\cite{GortonJohnson2024,Johnson2025PANASh}. In parallel, quantum-simulation studies have reduced qubit and gate requirements through compact encodings, symmetry projection and scalable Hamiltonian input schemes~\cite{SiwachLacroix2021,Singh2025,DuVary2025,Du2025abinitio}. In particular, symmetry-adapted Hamiltonians, drastically reduced in size by the near Sp(3,$\mathbb{R})\supset$ \SU{3} symmetry of nuclei~\cite{LauneyMD_ARNPS21}, have been shown to fit on current quantum hardware and to reproduce the $^{12}$C rotational band~\cite{mumma2024efficaciousband}.

In the entanglement studies mentioned above, the measures are evaluated from single-particle Slater-determinant expansions. In this paper, we obtain them from symmetry-guided large-scale calculations feasible beyond the lightest systems~\cite{LauneyMD_ARNPS21,launey2025_SAreview}, and use, to our knowledge for the first time, the SU(3)-coupling scheme to assess quantum-simulation resources at the level of two-body channels. This extends the resource reduction of Ref.~\cite{mumma2024efficaciousband} from the many-body basis to the excitation operators. We compare the resulting symmetry-channel and operator costs with those of the $jj$ scheme.

Entanglement structure also guides entanglement forging on quantum hardware~\cite{PerezObiol2026}, and has shaped scalable state-preparation strategies for nuclear ground states~\cite{Gu2026}. Turning entanglement information into a practical simulation strategy requires more than identifying strongly correlated orbitals. For example, orbital mutual information can guide orbital ordering, as demonstrated in density-matrix renormalization group (DMRG) calculations~\cite{Rissler2006}. Operator selection, in turn, is addressed by adaptive variational algorithms that construct an ansatz from a pool of candidate operators~\cite{Grimsley2019}. Mutual information has also been used to rank and screen candidate entanglers, reducing the pool used in adaptive variational calculations~\cite{Zhang2021}. Applications to nuclear structure include ADAPT-VQE calculations in the nuclear shell model~\cite{Romero2022,PerezObiol2023Sci} and studies of how excitation-operator ordering affects convergence in unitary coupled-cluster calculations of ${}^{6}\mathrm{Li}$~\cite{Kiss2022}. Connecting orbital correlations with candidate excitation operators therefore offers a way to use entanglement information in the design of quantum simulations.

In this paper, we start by building upon the work of Ref.~\cite{Robin2021}, which computes single-orbital entropies, in addition to two-orbital
mutual informations and negativities for $^4$He and $^6$He, and uses them to compare
harmonic-oscillator, Hartree-Fock (HF), natural, and variational natural single-particle bases, for use in quantum simulations. Here, we use the fact that mode-resolved entanglement measures can be evaluated from reduced density matrices whose matrix elements are exact contractions of the symmetry-coupled densities against state-independent unitary matrices, so they are available at scales where a Slater-determinant expansion is not practical, through the use of the \textit{ab initio} symmetry-adapted no-core shell model (SA-NCSM)~\cite{LauneyMD_ARNPS21,launey2025_SAreview}. In addition, we develop a second, distinct type of measure driven by symmetry irreducible representations (irreps).The two types of measures answer different questions, with mode entropies informing the register in a quantum simulation, while the symmetry-guided channels inform the circuit design.

Specifically, Sec.~\ref{sec:modes} works with single-particle modes $\ama$. Fock space factorizes exactly over modes, $ \bigotimes_{\ama} \mathbf{H}_{\ama}$~\cite{Zanardi2001}, so reduced density matrices come from fermionic partial traces and the entropies are entanglement entropies between modes. Under a Jordan-Wigner occupation encoding~\cite{JordanWigner1928}, one mode is encoded in one qubit, so these quantities speak directly to register layout, including qubit count, orbital ordering, and  which modes can be frozen. We also discuss the mode-summed linear (Tsallis) entropy, which is basis-independent and hence measures correlations that persists in every single-particle basis (Sec.~\ref{sec:subshell}). It brackets the natural-orbital entropy without diagonalizing the one-body density, while informing state preparation.

Sec.~\ref{sec:channels} works with pair operators for identical nucleons coupled to good SU(3) and spin quantum numbers. There is no mode partition and no partial trace. The objects are two-body density matrices~\cite{Coleman1963}, which give the pair occupations. These quantities inform the circuit design, e.g., which excitation operators carry weight and are hence selected as candidates for a variational operator pool or are kept in a truncation. We show that, in all cases considered, the dominant like-particle correlations are concentrated in fewer SU(3) channels than angular-momentum channels, while the number of operator components within the channels is comparable in the two schemes for the heavier nuclei. This points to an advantage of the SU(3)-coupling scheme in the number of variational parameters. In Sec.~\ref{sec:layout}, the retained channels also define correlation graphs used to construct the quantum register ordering.

\section{Theoretical framework}
\label{sec:defs}

We consider two types of entropies, and their difference is physically important. The Shannon entropy of a probability distribution $\{p_i\}$ (where $p_i \ge 0$ and $\sum_i p_i = 1$) is:
\begin{equation}
\Hb(\{p_i\}) = -\sum_i p_i \ln p_i ,
\label{eq:su3shannon}
\end{equation}
which expresses the uncertainty about which outcome occurs. Its two-outcome case with probabilities $h$ and $1-h$ is the binary entropy:
\begin{equation}
    \Hb(h) \equiv \Hb(\{h, 1-h\}) = -h\ln h - (1-h)\ln(1-h).
    \label{eq:su3Hb}
\end{equation}
The von Neumann entropy of a density matrix $\rho$ (Hermitian, positive semidefinite, and unit trace) is:
\begin{equation}
\mathcal{S}(\rho) = -\operatorname{Tr}\rho\ln\rho = -\sum_k \omega_k \ln\omega_k ,
\label{eq:su3vn}
\end{equation}
the Shannon entropy of the spectrum $\{\omega_k\}$ of $\rho$.
It is invariant under any unitary transformation $\rho \to U\rho U^\dagger$ of the space on which $\rho$ acts (a change of the single-particle basis is not of this kind for a reduced density matrix, since it changes the modes being traced over, and hence $\rho$ itself). In contrast, the Shannon entropy of the diagonal entries $\rho_{ii}$
depends on the basis. In any basis~\cite{Baumgratz2014}, we have:
\begin{equation}
\Hb(\{\rho_{ii}\}) \;\ge\; \mathcal{S}(\rho),
\qquad
C(\rho) \equiv \Hb(\{\rho_{ii}\}) - \mathcal{S}(\rho) \;\ge\; 0,
\label{eq:su3coherence}
\end{equation}
where the equality holds if and only if $\rho$ is diagonal in that basis. The gap $C(\rho)$ is the relative entropy of coherence, that is,  the part of the uncertainty due to superpositions of the basis states rather than to a classical mixture of them. Both entropies lie in $[0, \ln d]$ for a $d$-dimensional space and are quoted in nats, rather than bits.

In this study, mode entropies are calculated through the reduced density matrices in the Fock space of modes (filled or empty), which are in turn computed from nuclear one-body (OBD) and two-body densities (TBD) that are coupled in SU(3)$\times $SU(2), which are discussed next.

We use the shell label $a = (\eta_a\,0) \frac12$ with 
degeneracy $\Oms_a = \Pi^2_\frac12 \dim(\eta_a,0)$, where: 
\begin{equation}
\mathrm{dim}(\lambda,\mu) = (\lambda+1)(\mu+1)(\lambda+\mu+2)/2,
    \label{eq:su3dim}
\end{equation}
and $\Pi_J \equiv \sqrt{2J+1}$ for  angular momentum $J$.
Within it, the sublabel is
$m_a = (l_a m_{l_a} m_{s_a})$. 
The total dimension of
the model space is: 
\begin{equation}
\Om = \sum_a \Oms_a .
\label{eq:su3Omega}
\end{equation}
summed over all $a$ shells.  

The coupled pair-creation operator is:
\begin{equation}
P^{(\lambda\,\mu)S\alpha M\dagger}_{ab} = (1+\delta_{ab})^{-1/2}
\{ a^\dagger_{a} \times  a^\dagger_{b}\}^{(\lambda\,\mu)S\alpha M},
\label{eq:su3pairop}
\end{equation}
with spin rank $S$, orbital rank $L$, total angular momentum $J$, and conjugate operator:
\begin{equation}
  \tilde P^{(\lambda\,\mu)S\alpha M}_{ab} = (-1)^{\lambda + \mu + J + M} P^{\tilde{(\mu\,\lambda)}\kappa (LS)J,-M}_{ab},
\end{equation}
defined in terms of the SU(3) conjugate annihilation operator, $
  \tilde a_{\tilde a m_a} = (-1)^{\eta_a + l_a + m_{l_a} + \frac12 + m_{s_a}}\, a_{a,\,-m_{l_a},\,-m_{s_a}}$.

For a channel defined in terms of the SU(3)$\times$SU(2) irreps as:
\begin{equation}
c=(\lambda\mu)\,\kappa\,L\,S\,J\,M \equiv (\lambda\mu) S \alpha M,
\label{eq:su3chan}
\end{equation}
where $\alpha=\kappa L J$ with $\kappa$ labeling repeated $L$ in $(\lambda\mu)$,
 the coupled one-body operator:
\begin{equation}
  T^{(\lambda\,\mu)S\alpha M}_{ab} = \{ a^\dagger_a \times \tilde a_b \}^{(\lambda\,\mu)S\alpha M}  
  \label{Tph}
\end{equation}
and the coupled two-body operator: 
\begin{equation}
    T^{(\lambda\,\mu)S\alpha,\,(\lambda'\,\mu')S'\alpha';\,KM}_{ab,a'b'} =\big\{ P^{(\lambda\,\mu)S\alpha\dagger}_{ab} \times \tilde P^{(\lambda'\,\mu')S'\alpha'}_{a'b'}
\big\}^{KM}
\end{equation}
define the OBD and TBD reduced matrix elements as:
\begin{equation}
O^{(\lambda\,\mu)S\alpha}_{ab} \;=\; \avg{\Psi^{J_0} \,\big\|\, T^{(\lambda\,\mu)S\alpha}_{ab} \,\big\|\, \Psi^{J_0}}
\label{eq:su3su3obdme}
\end{equation}
\begin{equation}
O^{(\lambda\,\mu)S\alpha,\,(\lambda'\,\mu')S'\alpha';\,K}_{ab,\,a'b'} \;=\;
\avg{\Psi^{J_0} \,\big\|\, T^{(\lambda\,\mu)S\alpha,\,(\lambda'\,\mu')S'\alpha';\,K}_{ab,a'b'}  \,\big\|\, \Psi^{J_0}}.
\label{eq:su3su3tbdme}
\end{equation}
The $M_0$-averaged densities are defined as:
\begin{eqnarray}
&&\bar O^{(\lambda\,\mu)S\alpha M}_{ab}=\frac{1}{2J_0+1}\sum_{M_0}\bra{\Psi^{J_0 M_0}}\, T^{(\lambda\,\mu)S\alpha M}_{ab} \,\ket{\Psi^{J_0 M_0}} \cr
&=& \delta_{J0}\,\frac{O^{(LS)0}_{ab}}{\Pi_{J_0}},
\label{eq:su3Obar}\\[14pt]
&&\bar O^{(\lambda\,\mu)S\alpha M,\,(\lambda'\, \mu')S'\alpha'M}_{ab,\,a'b'}\cr
&=&\frac{1}{2J_0+1}\sum_{M_0}\bra{\Psi^{J_0 M_0}}\, P^{(\lambda\,\mu)S\alpha M\dagger}_{ab}\, P^{(\lambda'\,\mu')S'\alpha'M}_{a'b'} \,\ket{\Psi^{J_0 M_0}}
\cr
&=& \delta_{JJ'}\,\frac{O^{(\lambda\,\mu)S\alpha,\,(\lambda'\,\mu')S'\alpha';\,0}_{ab,\,a'b'}}{\Pi_J\,\Pi_{J_0}},
\label{eq:su3Obar2}
\end{eqnarray}
using 
$\sum_{M_0}\langle J_0M_0\,K0|J_0M_0\rangle = (2J_0+1)\,\delta_{K0}$ and $(-1)^{J-M}\langle JM\,J{-M}|00\rangle = 1/\Pi_J$. For $J_0 = 0$, these coincide with the $M_0$-projected expectation values of the one-body (OBDME) and two-body density matrix elements (TBDME).

\section{Symmetry-adapted basis}
\label{sec:SAbasis}

The SA-NCSM approach can accommodate larger model spaces, and  reach heavier nuclei than traditional methods, including $^{20}$Ne \cite{DytrychLDRWRBB20,DreyfussLESBDD20}, $^{21}$Mg \cite{Ruotsalainen19}, $^{22}$Mg \cite{Henderson:2017dqc}, $^{28}$Mg \cite{PhysRevC.100.014322}, as well as $^{32}$Ne and $^{48}$Ti \cite{LauneySOTANCP42018,LauneyMD_ARNPS21}, and $^{40}$Ca \cite{burrows2025responsemodel}. The SA approach with continuum, applicable to structure \cite{LauneyDD16,DytrychLDRWRBB20} and reactions \cite{MercenneLEDP19,DreyfussLESBDD20,LauneyMD_ARNPS21}, exploits new concepts that take maximum advantage of  the existence of symmetries of the nuclear many-body dynamics. In doing so, it resolves the scale explosion problem in nuclear structure calculations, {\em i.e.}, the explosive growth in computational resource demands with increasing number of particles and size of the spaces in which they reside. Achieving large model spaces is essential to accommodate important collective modes and to couple to continuum degrees of freedom within the interaction range, whereas beyond this, continuum Coulomb exact eigenfunctions are used. The aim is to solve this problem with unprecedented accuracy and beyond few-nucleon systems, while  providing nuclear structure and reaction information of interest to theoretical and experimental studies.

The many-nucleon basis states of the SA-NCSM are constructed using an efficient group-theoretical algorithm based on \SU{3}$\times$\SU{2}$_S$ configurations (irreducible representations, irreps) labeled by $(\lambda\,\mu)$ quantum numbers and intrinsic spin $S$. Hence, the basis states bring forward important information about nuclear deformation, according to an established  mapping \cite{CastanosDL88,RosensteelR77,LeschberD87}. For example, $(0 0)$, $(\lambda\, 0)$ and $(0\,\mu)$ describe spherical, prolate and oblate deformation, respectively. 

In particular, for all possible distributions of $A_p$ protons $\{A_{p_0}, A_{p_1}, A_{p_2}, \dots\}$ and $A_n$ neutrons $\{A_{n_0}, A_{n_1}, A_{n_2}, \dots\}$ over the major HO shells $\eta$ ($\eta=0,1,2,\dots$ for the $s$, $p$, $sd$,\dots HO shell), limited by the number of HO quantum excitations up through $\Nmax$, the \SU{3}$_\eta\times$\SU{2}$_{S_\eta}$ configurations are first enumerated for every major HO shell and separately for protons and neutrons:
\begin{widetext}
\begin{equation} 
\left|
\begin{array}{l}
\left(\left( \left( \gamma_{p_0}, \gamma_{p_1} \right)w'_p,\gamma_{p_2}  \right)w''_p,  \dots \right)\rho_p(\lambda_{p}\,\mu_{p})S_{p};
\\
\left(\left( \left( \gamma_{n_0}, \gamma_{n_1} \right)w'_n,\gamma_{n_2}  \right)w''_n,  \dots \right)\rho_n(\lambda_{n}\,\mu_{n})S_{n};
\end{array}
\rho(\lambda\,\mu)\kappa (LS) J M\right>, {\rm where } 
\begin{array}{ll}
\gamma_\eta =A_\eta \alpha_\eta (\lambda_{\eta}\,\mu_{\eta})S_{\eta} & \textrm{ (in-shell)}
\\
w= \rho (\lambda\,\mu)S & \textrm{ (inter-shell)},
\end{array}
 \label{SAbasis} 
\end{equation}
\end{widetext}
with $\gamma_p$ and $w_p$ ($\gamma_n$ and $w_n$) labeling protons (neutrons), as well as  $S_{p}$ and $S_{n}$ denoting proton and neutron intrinsic spins. The additional quantum numbers,   $\alpha_\eta$  (as described below) and $\rho$ are multiplicity labels. The resulting proton, $(\lambda_{p}\,\mu_{p})S_{p}$,  and neutron, $(\lambda_{n}\,\mu_{n})S_{n}$,  configurations are coupled to good $(\lambda\,\mu)$ and total intrinsic spin $S$. The basis states are then classified according to $\SU{3}   \stackrel{ \kappa}{\supset}\SO{3}_L$, where the label $\kappa$ distinguishes multiple occurrences of the same orbital angular momentum $L$ in
the parent irrep $(\lambda\,\mu)$. Similar to the $LS$-coupling scheme, $L$ is coupled with $S$ to the total angular momentum $J$ and its projection $M$. In this way, a complete  shell-model basis is classified.

An important feature of the \SU{3}-coupling scheme is that  all possible \SU{3}$_\eta\times$\SU{2}$_{S_\eta}$ configurations  within a major HO shell $\eta$ (for protons or neutrons) are readily available based on  the $U(\Omega_\eta)$ unitary group of the many-body three-dimensional HO, where $\Omega_{\eta}=(\eta+1)(\eta+2)/2$ is the number of distinguishable distributions of $\eta$ quanta in the three Cartesian directions, $x$, $y$, and $z$ (spatial degeneracy), e.g., for $\eta=2$, there are 6 different distributions, $(\eta_x,\eta_y,\eta_z)=(2,0,0),(1,1,0),(1,0,1),(0,2,0),(0,1,1)$ and $(0,0,2)$. In particular, the basis construction is implemented according to the reduction \cite{DraayerLPL89}:
\begin{equation}
\begin{array}{cccccc}
   \Un{\Omega_\eta}        &     & & \times  &  \SU{2}       \\
          \left[f_1,f_2,\dots f_{\Omega_\eta}\right] &     & & & S_\eta       \\
  \cup   &  \alpha_\eta   & &    &     \\
\SU{3}   & &   & &     \\
(\lambda_{\eta}\,\mu_{\eta})   & &   & &     \\
\end{array},
\end{equation}
where a multiplicity index $\alpha_{\eta}$ distinguishes multiple occurrences of an
\SU{3} irrep $(\lambda_{\eta}\,\mu_{\eta})$ in a given $\Un{\Omega_{\eta}}$ irrep labeled by Young tableaux, $[\mathbf{f}]=[f_1,f_2,\dots,f_{\Omega_\eta}]$,  with $f_i=0,1,2$, for all $i$  -- this corresponds to the number of particles (up to two) occupying a given  $(\eta_x,\eta_y,\eta_z)$ configuration~\cite{LauneyDD16}.  Using  $\SpR{3} {\supset} \SU{3}   \stackrel{ \kappa}{\supset}\SO{3}_L$, the \SU{3}-adapted basis can be further organized into \SpR{3}-preserving subspaces, after diagonalization of the second-order \SpR{3} Casimir invariant, as done in Ref.~\cite{DytrychLDRWRBB20}. 

In this work, we perform all nuclear structure calculations in the SA-NCSM framework with NNLO$_{\rm opt}$ \cite{Ekstrom13} in $\Nmax=12$ for $^{4,6}$He and $\Nmax=8$ for $^8$He in complete model spaces, and $\Nmax=16$ for $^8$Be, $\Nmax=8$ for $^{12}$C {($\Nmax=6$ in Fig.~\ref{fig:8Be_su3_sp3r})}, $\Nmax=6$ for $^{20}$Ne and $^{48}$Ca in selected model spaces, using $\hbar\Omega = 15$ MeV for all nuclei except $^6$He, which uses $\hw=16$ MeV.
We have shown that these selected model spaces retain the necessary SU(3) configurations in the many-body wavefunction that reproduce observables of the corresponding complete $\Nmax$ spaces~\cite{LauneyDD16,DytrychLDRWRBB20,DreyfussLESBDD20,LauneyMD_ARNPS21}. Selected spaces of this kind have been also used in the 
quantum simulation for $^{12}$C of Ref.~\cite{mumma2024efficaciousband}.

\subsection{Entropies for symmetry-adapted bases}

\begin{figure}[H]
\centering
\includegraphics[width=1.0\linewidth]{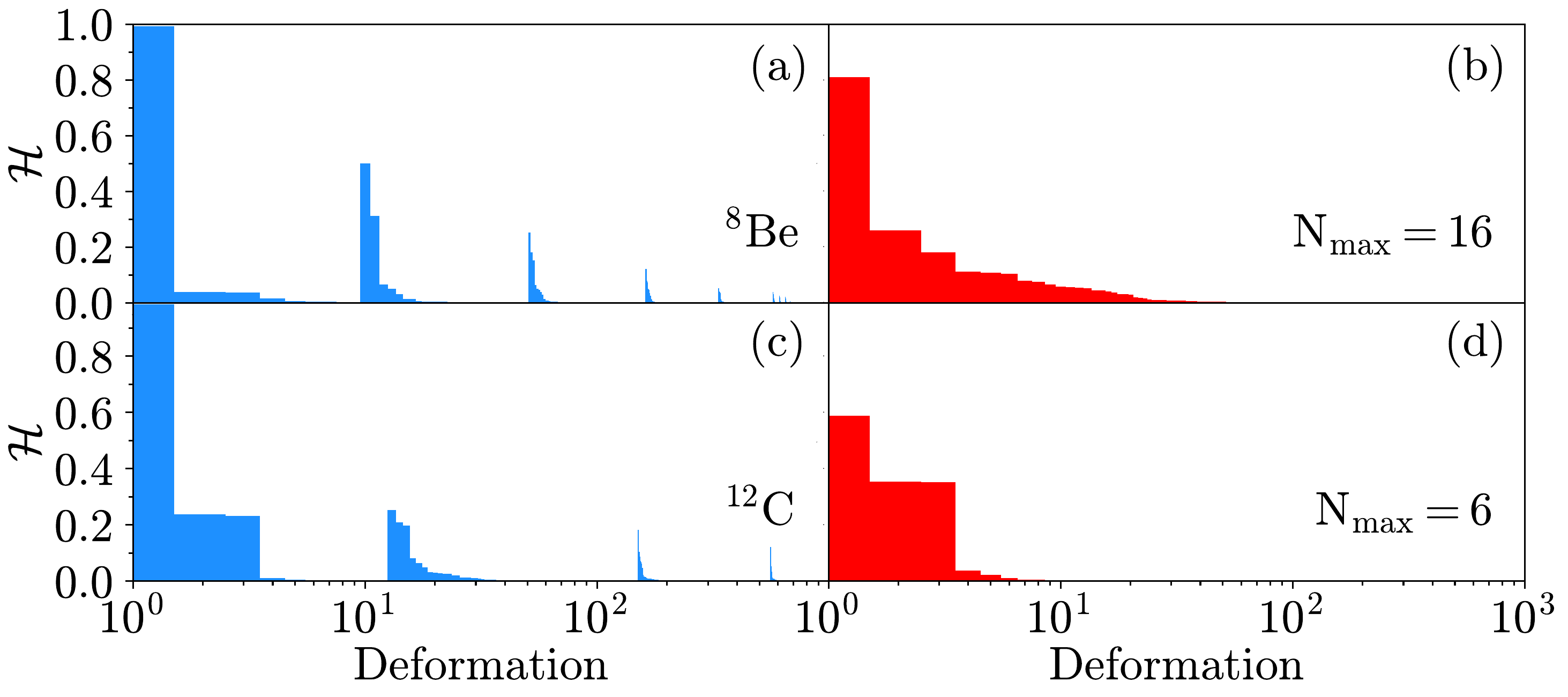}
\vspace{-0.5cm}\caption{
Shannon entropy $\mathcal{H}$ of the distribution of the ground state of $^8$Be and $^{12}$C over the SU(3) (blue bars) and \SpR{3} (red bars) symmetry-adapted bases states. Each bar gives the contribution of one irrep, plotted against its    second-order SU(3) Casimir invariant value that inform its deformation, in model spaces given by the $\Nmax$ values shown and $\hw=15$ MeV. The uneven width of the bars is due to the log axis and bears no physical significance. The dominant bars in panels (a)\&(c) are 0.99 nats, but are visually indistinct from 1 nat.
}
\label{fig:8Be_su3_sp3r}
\end{figure}

The SU(3) and \SpR{3} symmetry-adapted bases distribute the same ground state very differently (Fig.~\ref{fig:8Be_su3_sp3r}). We use the Shannon entropy~\eqref{eq:su3shannon} for the probability amplitudes over all basis states, and resolve it by irrep, so that each bar is the contribution of the basis states of one SU(3) or \SpR{3} irrep. In the SU(3) basis, this entropy is spread over well-defined groups of irreps, as discussed below [panels (a) and (c)]. In the Sp(3,$\mathbb{R}$) basis, it is confined to a few irreps, each of which with a bandhead of low deformation (that is, in low $\hw$ subspaces). We note that this irrep contains the highly deformed particle-hole excitations, and steadily decreases for bandheads in high $\hw$ subspaces [panels (b) and (d)]. 

The difference reflects the structure of Sp(3,$\mathbb{R}$) basis configurations: each Sp(3,$\mathbb{R}$) irrep contains a bandhead, which is an SU(3) irrep, together with the $2\hbar\Omega$, $4\hbar\Omega$,
\dots\ particle-hole monopole and quadrupole excitations that increase the deformation, so that the high-deformation SU(3) groups belong to the few Sp(3,$\mathbb{R}$) irreps~\cite{LauneyDD16,DytrychLDRWRBB20}.
In $^{8}$Be a single Sp(3,$\mathbb{R}$) irrep carries most of the entropy, whereas in $^{12}$C several low-lying irreps contribute comparably. Since $e^{\Hb}$ is the effective number of basis states occupied by the ground state,
the Sp(3,$\mathbb{R}$) organization requires fewer basis states and allows a truncation at a single deformation with little loss, which underpins the reduction of the many-body basis in Ref.~\cite{mumma2024efficaciousband}. In the sections below, we discuss whether the same symmetry also concentrates the two-body correlations that set the operator content of a quantum circuit.

\section{Mode entropy: informing the register}
\label{sec:modes}

\subsection{One-body density matrix elements for orbits (modes)}

Mode entropies are defined in terms of the (uncoupled) one-body density matrix elements as:
\begin{equation}
\gamma_{\ama,\,\bmb} = \avg{\Psi^{J_0M_0} \,|\, a^\dagger_{\ama}\, a_{\bmb} \,|\, \Psi^{J_0M_0}}
\label{eq:su3gammadef}
\end{equation}
with $\Oms_a \times \Oms_b$ matrix blocks  $
\big[\gamma_{ab}\big]_{m_a m_b} = \gamma_{\ama,\, \bmb}$, and are calculated using the SU(3)-coupled densities in Eq.~\eqref{eq:su3invert}. This coupling naturally results in the single-particle mode $\ama=\eta_a l_a m_{l_a} m_{s_a}$. We note that a trivial unitary transformation can change $\ama$ from $(\eta_a l_a m_{l_a} m_{s_a})$ to $(\eta_a l_a j_a m_{a})$. We refer to these as spin-orbital modes or $j$-modes, respectively.  Hence, the $\gamma_{j_am_{j_a},j_bm_{j_b}}$ matrix for $j$-modes, as used in Ref.~\cite{Robin2021}, can be straightforwardly calculated and is used in the paper when the $j$-modes are discussed.

While the off-diagonal blocks carry the inter-subshell coherences, each diagonal block $\gamma_{aa}$ is a principal submatrix, Hermitian and positive semidefinite with eigenvalues in $[0,1]$, with diagonal elements:
\begin{equation}
\gamma_{\ama} \equiv \gamma_{\ama,\,\ama}= \avg{\Psi^{J_0M_0} \,|\, \hat N_{\ama} \,|\, \Psi^{J_0M_0}},
\label{eq:su3gammadiagdef}
\end{equation}
that are the expectation value of the mode number operator, expressed in terms of Eqns.~\eqref{Tph} and \eqref{eq:su3U}:
\begin{equation}
  \hat N_{\ama} = a^\dagger_{\ama} a_{\ama}  = \sum_{\lambda S\alpha} \big[U^{(\lambda\,\lambda)S\alpha0}_{aa}\big]_{m_a m_a}\, T^{(\lambda\,\lambda)S\alpha0}_{aa}.
\label{eq:su3NT}
\end{equation}
Here, for a general wavefunction with $J_0$ ($J \le 2J_0$), only two sets of quantum numbers contribute, since  $S=0$ or $1$:
\begin{equation}
(\lambda\,\lambda)(L,0)\,J = L \qquad\text{and}\qquad (\lambda\,\lambda)(L,1)\,J = L \pm 1,
\label{eq:su3families}
\end{equation}

The subshell occupation $n_a=\avg{\Psi^{J_0M_0} \,|\, \hat N_a \,|\, \Psi^{J_0M_0}}=\avg{\hat N_a}$ is given by the subshell number operator
$\hat N_a = \sum_{m_a} \hat N_{\ama} = \sqrt{\dim_a} T^{(00)\kappa=0(LS=00)JM=00}_{aa}\equiv T^{0}_{aa}$, which is
invariant under deformation $(\lambda\,\mu)$, orbital, and spin rotations separately [we will use ``$0$" to denote the scalar channel $c=0$, given by the set ${(\lambda \, \mu)=(00),\kappa=0,(LS=00),JM=00}$]. Hence, we have:
\begin{equation}
n_a \;=\; \sum_{m_a} \gamma_{\ama} \;=\; \operatorname{Tr}\gamma_{aa}
 \;\in\; [0,\, \Oms_a],
\qquad \sum_a n_a = A ,
\label{eq:su3na1}
\end{equation}
where $A$ is the total number of nucleons. We note that if proton (neutron) densities are used, $A$ will denote the number of protons (neutron).
Using Eq.~\eqref{eq:su3trU} then yields:
\begin{eqnarray}
    n_a = \operatorname{Tr}\gamma_{aa} &=& \sum_{(\lambda\,\mu)S\alpha} \bar O^{(\lambda\,\mu)S\alpha0}_{aa}\, \operatorname{Tr}\, U^{(\lambda\,\mu)S\alpha0}_{aa}  \cr &=& \sqrt{\Oms_a}\;\bar O^{0}_{aa}.
    \label{eq:su3na2}
\end{eqnarray}
Hence, only the scalar channel contributes to $n_a$, and all other non-scalar $U_{aa}^{(\lambda\,\mu)S\alpha0}$ are traceless. 

For $J_0 = 0$, only $(\lambda \, \mu) (LS)J=(\lambda\,\lambda) (00)0$ and $(\lambda \, \lambda) (11)0$ remain according to Eq.~\eqref{eq:su3families}, and $\gamma_{\ama}$  becomes:
\begin{equation}
\begin{aligned}
\gamma_{\ama}&=\frac{n_a}{\Oms_a}
+\frac{1}{\Pi_{l_{a}}\Pi_{1/2}}\sum_{\lambda=2,4,\dots}c^{(0)}_{{l_{a}}\lambda}\,\bar O^{(\lambda\, \lambda)\,(00)\,0}_{aa}\\
&\quad-\frac{\sqrt2\;m_{l_{a}}\,m_s}{\sqrt{{l_{a}}({l_{a}}+1)(2{l_{a}}+1)}}\sum_{\lambda\ \text{odd}}c^{(1)}_{{l_{a}}\lambda}\,
\bar O^{(\lambda\, \lambda)\, (1,1)\,0}_{aa},
\end{aligned}
\label{eq:su3xJ0}
\end{equation}
where we remind that $\ama=\eta_a l_a m_{l_a} m_{s_a}$ and we use $c^{(L)}_{l\lambda}=(-1)^{\eta_a}\,\langle(\eta_a\,0)\,l_a;(0\,\eta_a)\,l_a\,\|\,(\lambda\, \lambda)\,L\rangle_1$, along with $\bar O^{(00)0(00)00}_{aa}U^{(00)0(00)00}_{aa} = \frac{n_a}{\sqrt{\Oms_a}} \frac{\one}{\sqrt{\dim_a}} $ from Eq.~\eqref{eq:su3na2}. 

\subsection{Single-mode reduced density}

The Fock space of a mode $\ama$ is $\{\ket{0}_{\ama}, \ket{1}_{\ama}\}$ (empty or filled). Its reduced density
matrix is given by the partial trace over the remaining modes~\cite{Robin2021}. It is determined through $\gamma_{\ama}$ of Eq.~\eqref{eq:su3gammadiagdef} calculated using Eq.~\eqref{eq:su3gamma}, with off-diagonal elements vanishing by
particle-number conservation, and for any $J_0$ is given as:
\begin{equation}
\rho^{(\ama)} = \begin{pmatrix} 1-\gamma_{\ama} & 0 \\ 0 & \gamma_{\ama} \end{pmatrix}.
\label{eq:su3rho1}
\end{equation}
The von Neumann entropy is given as:
\begin{equation}
\begin{gathered}
\mathcal{S}_{\ama} = -\operatorname{Tr}\left[\rho^{(\ama)} \ln \rho^{(\ama)}\right] = \Hb(\gamma_{\ama})\\
=-\sum_{k=1}^{2}\omega_{k}^{(\ama)}\ln \omega_{k}^{(\ama)},
\label{eq:su3rho1modeS}  
\end{gathered}
\end{equation}
where $\omega_{k}^{(\ama)}=\{1-\gamma_{\ama},\gamma_{\ama}\}$
are the eigenvalues of $\rho^{(\ama)}$ and $\Hb(h) $ is defined in Eq.~\eqref{eq:su3Hb}.
Because $\rho^{(\ama)}$ is
diagonal, its eigenvalues are its diagonal entries and the von Neumann entropy coincides with
the Shannon entropy of the occupation distribution $\{1-\gamma_{\ama}, \gamma_{\ama}\}$. For a
pure state, it gives the entanglement entropy between mode $\ama$ and the complementary Fock space. 

Diagonalizing the one-body densities mixes radial indices and gives natural occupations
$\tilde n^{(lj)}_i \in [0,1]$, with $i$ labeling the eigenvectors and $\sum_{l j i}(2j+1)\,\tilde n^{(lj)}_i = A$. Then the quantity:
\begin{equation}
\mathcal{S}_1 = \sum_{l\, j\, i} (2j+1)\, \Hb\big(\tilde n^{(lj)}_i\big) 
\label{eq:su3S1}
\end{equation}
is equal to $\sum_{\ama} \mathcal{S}_{\ama}$ in the
natural basis (used in Ref.~\cite{Robin2021}), or equivalently the correlation entropy of Ref.~\cite{Bulgac2023}, resolved by
$(l,j)$ with strict additivity.

\begin{figure}[ht]
    \centering
    \includegraphics[width=1\linewidth]{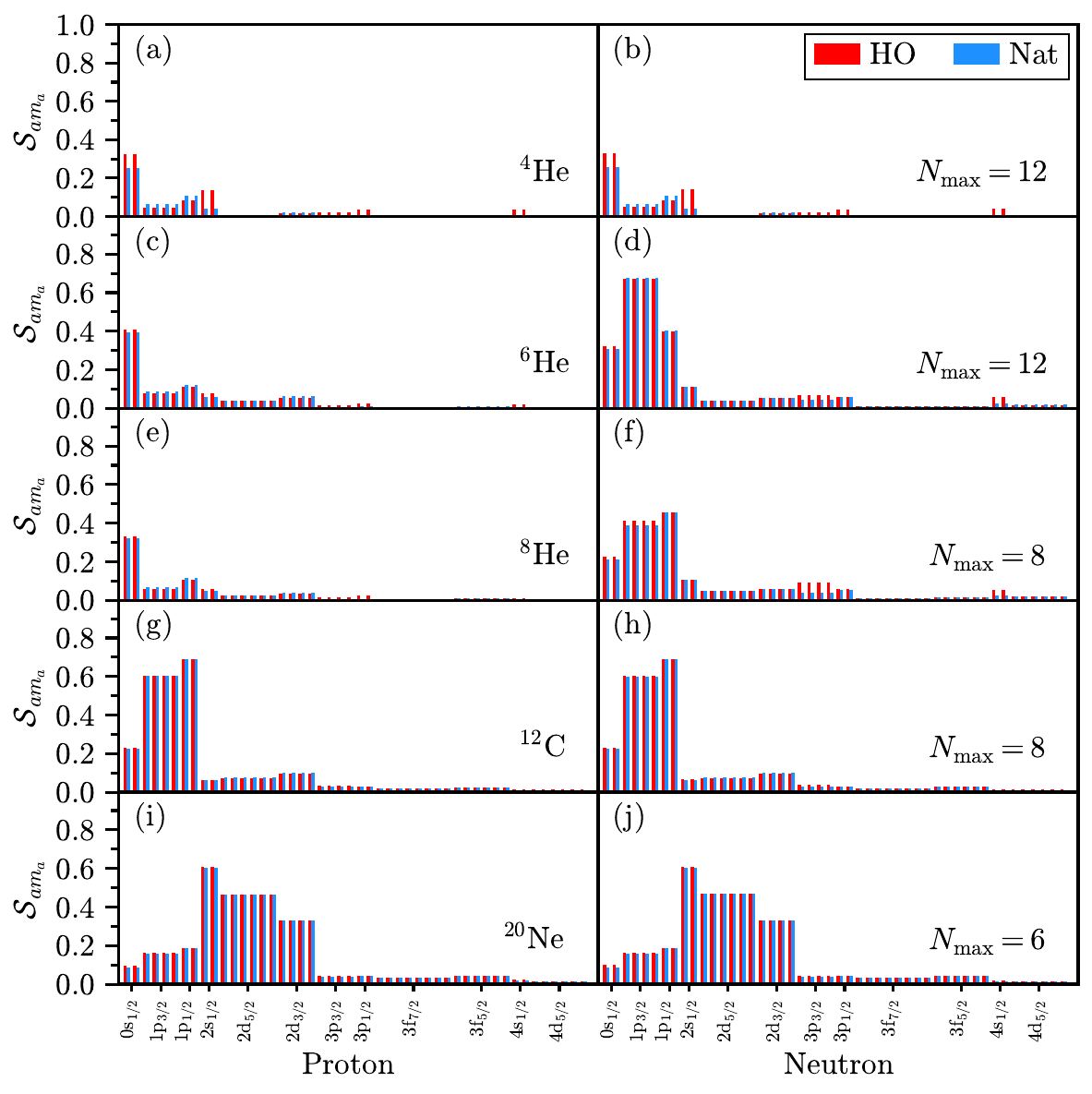}
    \vspace*{-0.5cm}\caption{
        Von Neumann entropy for harmonic oscillator (``HO'') and natural (``Nat'') orbits, 
        $j$ levels with physical ordering of the basis, and for the proton and neutron sectors of the nuclei $^{4,6,8}$He, $^{12}$C, and $^{20}$Ne. The $\Nmax$ value used for each nucleus is displayed in its respective panel and $\hw=15$ MeV is used for all nuclei except $^6$He, which uses $\hw=16$ MeV.
    }
    \label{fig:su2vonNeumannEntropy}
\end{figure}

We offer $\mathcal{S}_{am_a}$ for multiple nuclei and $\Nmax$ values to compare with Ref.~\cite{Robin2021} for $^{4,6}$He and investigate higher mass nuclei for $j$-level modes (see Fig.~\ref{fig:su2vonNeumannEntropy}).  
Overall, the total entropy grows with mass number.
We note that we use $\Nmax$ truncation to exactly remove c.m. spuriosity in the wavefunction, whereas Ref.~\cite{Robin2021} includes all possible configurations in the active single-particle space. These different truncation schemes lead, most notably, to small differences in $^4$He for the $1p_{\frac12}$ and $2s$ entropies. Similar to Ref.~\cite{Robin2021}, we find that entropy does not differ for equal numbers of protons and neutrons [see panels (a)\&(b), (g)\&(h), and (i)\&(j)]. Natural orbits necessarily have lower entropy summed over modes, and orbit by orbit, the reduction is largest in the $0s$ and $2s$ orbits of $^{4}$He [panels (a)\&(b)], and in the $3p_{\frac32}$ orbit for the neutrons of $^{6}$He [panel (d)]. Differences between natural and HO orbits largely decrease as one moves from the He isotopes toward $^{12}$C and $^{20}$Ne, which is clearest if one follows the trend in the left panels. This may be related to the $\hw$ used, which is very close to the optimal one, also called self-consistent, for these two nuclei, and is the one that yields the fastest convergence of observables, such as radii and quadrupole moments. In addition, we find that high-shell tails are more prominent and extend further for higher mass, but entropy is highly concentrated in the valence shell in all cases.

\subsection{Two-body density matrix elements for orbits (modes)}

The two-body density matrix elements that are needed to explore entanglement are defined as~\cite{Robin2021}:
\begin{equation}
\Gamma_{\ama,\,\bmb} = \avg{\Psi^{J_0M_0} | \hat N_{\ama}\, \hat N_{\bmb} | \Psi^{J_0M_0}}.
\label{eq:su3Gammadef}
\end{equation}
With the expression: 
\begin{equation}
\hat N_{\ama} \hat N_{\bmb}
= \delta_{ab}\delta_{m_a m_b}\, \hat N_{\ama}
+ a^\dagger_{\ama} a^\dagger_{\bmb} a_{\bmb} a_{\ama},
\label{eq:su3NNsplit}
\end{equation}
$\Gamma_{\ama,\bmb}$ can then be related to the coupled diagonal TBDME in Eq.~\eqref{eq:su3Obar2} (or pair terms):
\begin{equation}
\begin{aligned}
\Gamma_{\ama,\bmb} 
=& 
\sum_{\substack{(\lambda\,\mu)S\alpha\\ (\lambda'\, \mu')S'\alpha'}}
\big[U^{(\lambda\,\mu)S\alpha M,\,(\lambda'\, \mu')S'\alpha' M}_{ab}\big]_{m_a m_b}\\
\times\,&
\bar O^{(\lambda\,\mu)S\alpha M,\,(\lambda'\, \mu')S'\alpha' M}_{ab,\,ab},\quad \ama \ne \bmb.
\end{aligned}
\label{eq:su3pi}
\end{equation}
For coincident modes $\Gamma_{\ama,\ama} = \avg{\hat N_{\ama}^2} = \gamma_{\ama}$, which is given by the one-body density matrix. Hence, $\operatorname{Tr} \Gamma_{ab} = \sum_{m_a m_b} \Gamma_{\ama,\bmb} = \avg{\hat N_a \hat N_b}$, whereas $\avg{\hat N_a^2} = n_a + \sum_{m_a \ne m'_a} \Gamma_{\ama,\,a m'_a}$ for $a = b$. 

\subsection{Two-mode reduced density}

The two-mode reduced density matrix is written in the basis of joint occupations, ordered
$\ket{00}, \ket{01}, \ket{10}, \ket{11}$, with the first entry referring to mode $\ama$ and the
second to mode $\bmb$. Its corner $\bra{00}\rho^{(\ama,\bmb)}\ket{11}$ vanishes by
particle-number conservation alone, the off-diagonal element $\gamma_{\ama,\bmb}$ Eq.~\eqref{eq:su3gamma} is the
coherence, and $\Gamma_{\ama,\bmb}$ Eq.~\eqref{eq:su3pi} is the pair term. The full matrix is
given by:
\begin{widetext}
\begin{equation}
\rho^{(\ama,\bmb)} =
\begin{pmatrix}
1 - \gamma_{\ama} - \gamma_{\bmb} + \Gamma_{\ama,\bmb} & 0 & 0 & 0\\
0 & \gamma_{\bmb} - \Gamma_{\ama,\bmb} & \gamma^{*}_{\ama,\bmb} & 0\\
0 & \gamma_{\ama,\bmb} & \gamma_{\ama} - \Gamma_{\ama,\bmb} & 0\\
0 & 0 & 0 & \Gamma_{\ama,\bmb}
\end{pmatrix}.
\label{eq:su3rho2}
\end{equation}
\end{widetext}

\subsection{Entropy, mutual information, and negativity}
\label{sec:modemeasures}

The block structure of Eq.~\eqref{eq:su3rho2} gives the spectrum in closed form:
\begin{equation}
\begin{aligned}
p_{00} &= 1 - \gamma_{\ama} - \gamma_{\bmb} + \Gamma_{\ama,\bmb},\\
p_{11} &= \Gamma_{\ama,\bmb},\\
p_\pm &= \tfrac12\big(\gamma_{\ama} + \gamma_{\bmb} - 2\Gamma_{\ama,\bmb}\big)\\
&\quad \pm \sqrt{\tfrac14\big(\gamma_{\ama} - \gamma_{\bmb}\big)^2
+ |\gamma_{\ama,\bmb}|^2},
\end{aligned}
\label{eq:su3eigs}
\end{equation}
where $p_{00}$ and $p_{11}$ are the populations of $\ket{00}$ and $\ket{11}$ and $p_\pm$ diagonalize the $\{\ket{01},\ket{10}\}$ block. The two-mode von Neumann entropy is
$$\mathcal{S}_{\ama,\bmb} = -\operatorname{Tr}\rho^{(\ama,\bmb)}\ln\rho^{(\ama,\bmb)} = -\sum p\ln p$$
over these four eigenvalues, and quantum mutual information between the two modes is
$$I_{\ama,\bmb} = [\mathcal{S}_{\ama} + \mathcal{S}_{\bmb} - \mathcal{S}_{\ama,\bmb}](1-\delta_{\ama,\bmb}),$$
where the $\delta$-factor is introduced to ensure vanishing correlation between a single orbital and itself. Unlike the one-mode case, the eigenvalues $p_\pm$ differ from the diagonal entries $\gamma_{\bmb} - \Gamma_{\ama,\bmb}$ and $\gamma_{\ama} - \Gamma_{\ama,\bmb}$, so the Shannon entropy of the diagonal exceeds $\mathcal{S}_{\ama,\bmb}$ by the coherence $C(\rho^{(\ama,\bmb)})$ [Eq.~\eqref{eq:su3coherence}].

\begin{figure}[ht]
\centering
\includegraphics[width=1.0\linewidth]{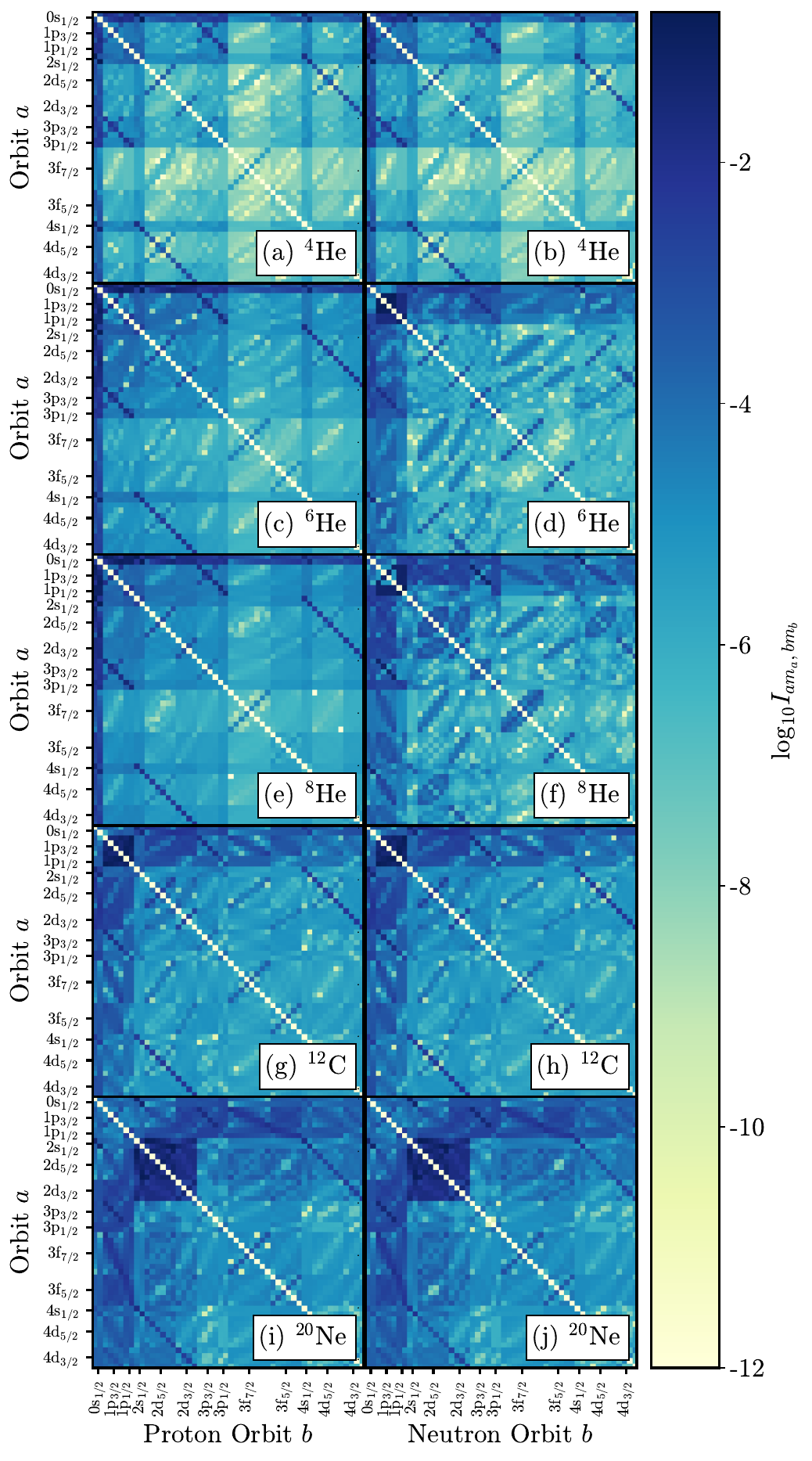}
\vspace*{-0.5cm}\caption{
Mutual information for harmonic oscillator orbits, $j$ levels, and the proton and neutron sectors of the nuclei $^{4,6,8}$He, $^{12}$C, and $^{20}$Ne. The $\Nmax$ and $\hw$ values are the same as in Fig.~\ref{fig:su2vonNeumannEntropy}.
}
\label{fig:su2MutualInfo}
\end{figure}
We offer $I_{\ama,\bmb}$ for multiple nuclei and $\Nmax$ values to compare with Ref.~\cite{Robin2021} for $^{4,6}$He and investigate higher mass nuclei for $j$-level modes (see Fig.~\ref{fig:su2MutualInfo}). The main diagonal is zero by construction, as described above in the definition of mutual information. The vertical and horizontal bands are a single orbit with several $m$-projections. The diagonals within a $j$-block are correlations between two orbits with aligned spin ($m_a=m_b$), whereas the antidiagonals within a band are correlations between two orbits with counter-aligned spins ($m_a=-m_b$). All fine structure is aligned along antidiagonals except between orbits with $l_a\,j_a\,m_a = l_b\,j_b\,m_b$, which is most visible in the heavier nuclei [see panels (g)-(j)]. This exception follows from the rotational invariance, namely, for the $J_0=0$ states studied here, rotational and parity symmetry restrict nonzero off-diagonal one-body coherences in the $j$-mode basis to radial partners with the same $(\ell,j,m)$. Similar to Ref.~\cite{Robin2021}, we find that the largest mutual information in $^4$He is between the $0s$ and $2s$ modes with the same $m$, and add that such radial-partner correlations remain significant in the heavier nuclei. As with von Neumann entropy, we also find that mutual information does not differ for equal numbers of protons and neutrons [see panels (a)\&(b), (g)\&(h), and (i)\&(j)]. For heavier nuclei than $^4$He [panels (d) and (f)-(j)], we see the highest contribution (darkest blue blocks) to the mutual information are confined within the valence shell.  

In addition, positivity of Eq.~\eqref{eq:su3eigs} gives bounds coupling one- and two-body data:
\begin{equation}
\begin{aligned}
&\max(0,\, \gamma_{\ama} + \gamma_{\bmb} - 1) \;\le\; \Gamma_{\ama,\bmb},\\[2pt]
&\Gamma_{\ama,\bmb} \;\le\; \min(\gamma_{\ama},\, \gamma_{\bmb}),\\[2pt]
&|\gamma_{\ama,\bmb}|^2 \;\le\;
(\gamma_{\ama} - \Gamma_{\ama,\bmb})(\gamma_{\bmb} - \Gamma_{\ama,\bmb}).
\end{aligned}
\label{eq:su3bounds}
\end{equation}

The negativity is given as \cite{Vidal2002}: 
\begin{equation}
\mathcal{N}^{(\ama,\bmb)}=\frac{\left\Vert \rho^{T_{\ama}(\ama,\bmb)}\right\Vert_1 -1}{2},\label{eq:su3Negativity}
\end{equation} 
which is equal to the sum of the absolute values of the negative eigenvalues of $\rho^{T_{\ama}(\ama,\bmb)}$, the partially-transposed two-mode reduced density matrix.
 
Partial transposition on mode $\bmb$ moves the coherence into the
$\{\ket{00},\ket{11}\}$ corner~\cite{Peres1996,Horodecki1996,Vidal2002}, giving:
\begin{equation}
\begin{aligned}
\mathcal{N}^{(\ama,\bmb)} &= \max\Big[0,\;
\sqrt{\tfrac14 (p_{00} - p_{11})^2 + |\gamma_{\ama,\bmb}|^2}\\
&\hspace{6.5em} - \tfrac12 (p_{00} + p_{11})\Big],\\[3pt]
\mathcal{N}^{(\ama,\bmb)} &> 0 \iff |\gamma_{\ama,\bmb}|^2 > p_{00}\, p_{11} .
\end{aligned}
\label{eq:su3neg}
\end{equation}
Two-mode negativity therefore requires nonzero coherence: if $\gamma_{\ama,\bmb} = 0$ the pair is separable and $I_{\ama,\bmb}$ is purely classical, however large. Combined with Eq.~\eqref{eq:su3bounds}, entangled mode pairs occupy the window
$p_{00}\, p_{11} < |\gamma_{\ama,\bmb}|^2 \le
(\gamma_{\ama} - \Gamma_{\ama,\bmb})(\gamma_{\bmb} - \Gamma_{\ama,\bmb})$. 

Because the pair of modes is correlated with the remainder of the nucleus, $\rho^{(\ama,\bmb)}$ is mixed. For a mixed state, $\mathcal{S}_{\ama,\bmb}$ reflects this correlation and is not a measure of correlation between $\ama$ and $\bmb$. That role is played by $I_{\ama,\bmb}$, which is the total (classical plus quantum) correlation between the two modes~\cite{Henderson2001}. A classically correlated state can have large $I_{\ama,\bmb}$ without any entanglement between them. The quantum part of $I_{\ama,\bmb}$ can be isolated with the two-orbital discord of Faba \emph{et al.}~\cite{Faba2021}, and entanglement proper means non-separability, i.e., \ $\rho^{(\ama,\bmb)}$ is not a mixture of product states~\cite{Werner1989}. For two modes, that is, two qubits, the Peres--Horodecki criterion is necessary and sufficient~\cite{Peres1996,Horodecki1996}, so Eq.~\eqref{eq:su3neg} decides entanglement exactly. This entanglement is, however, not accessible to operations local to each mode: for a single mode, local particle number and parity coincide, so under either superselection rule~\cite{Ding2021}, no such operation can create or detect superpositions of $|0\rangle$ and $|1\rangle$, and the coherence $\gamma_{\ama,\bmb}$ between $|01\rangle$ and $|10\rangle$ cannot be detected by local measurements. The negativity of Eq.~\eqref{eq:su3neg} is thus entanglement of the mode description rather than a resource for local operations. On a qubit register, where no superselection rule applies, it is entanglement that the circuit must create.
 
The two-mode matrix falls into one of three regimes:
\begin{center}
\setlength{\tabcolsep}{3.5pt}
\renewcommand{\arraystretch}{1.25}
\small
\begin{tabular}{p{0.34\columnwidth}p{0.60\columnwidth}}
\toprule
Coherence & Correlation\cr
\hline
\raggedright{$\gamma_{\ama,\bmb} = 0$} &
\raggedright{Classical only. $\rho^{(\ama,\bmb)}$ is diagonal in the product basis (classical-classical); $I$ is purely classical and the discord vanishes.}\cr
\raggedright{$0 < |\gamma_{\ama,\bmb}|^2 \le p_{00}p_{11}$} &
\raggedright{Quantum but separable. Generically nonzero discord~\cite{Ollivier2001,Faba2021}, no entanglement.}\cr
\raggedright{$|\gamma_{\ama,\bmb}|^2 > p_{00}p_{11}$} &
\raggedright{Entangled; $\mathcal{N}^{(\ama,\bmb)} > 0$.}\cr
\hline
\end{tabular}
\end{center}
The classical/quantum boundary is the coherence gap $C(\rho^{(\ama,\bmb)})$ of Eq.~\eqref{eq:su3coherence}, which vanishes exactly in the first regime. 

\begin{figure}
    \centering
    \includegraphics[width=0.85\linewidth]{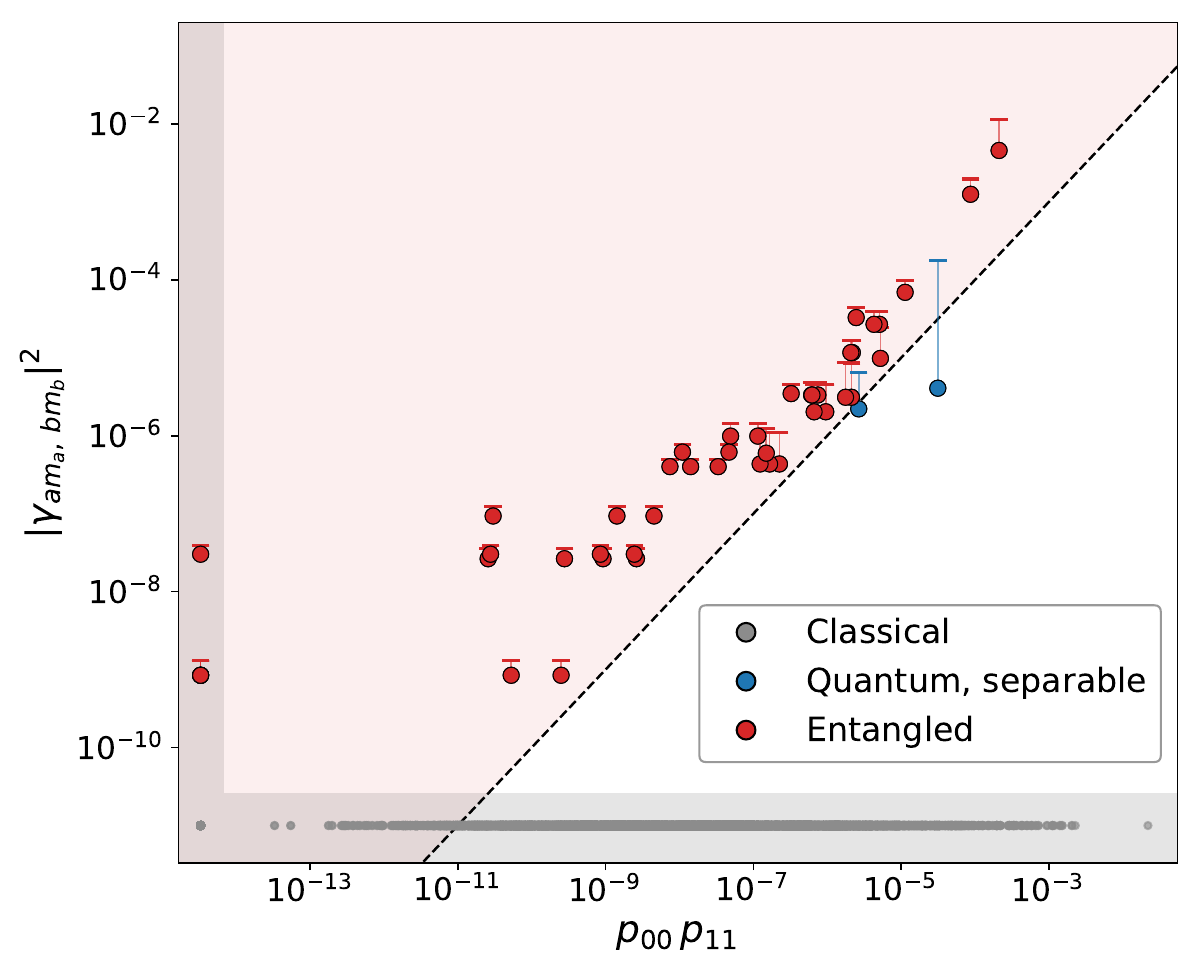}\\
    (a)\\
    \includegraphics[width=0.85\linewidth]{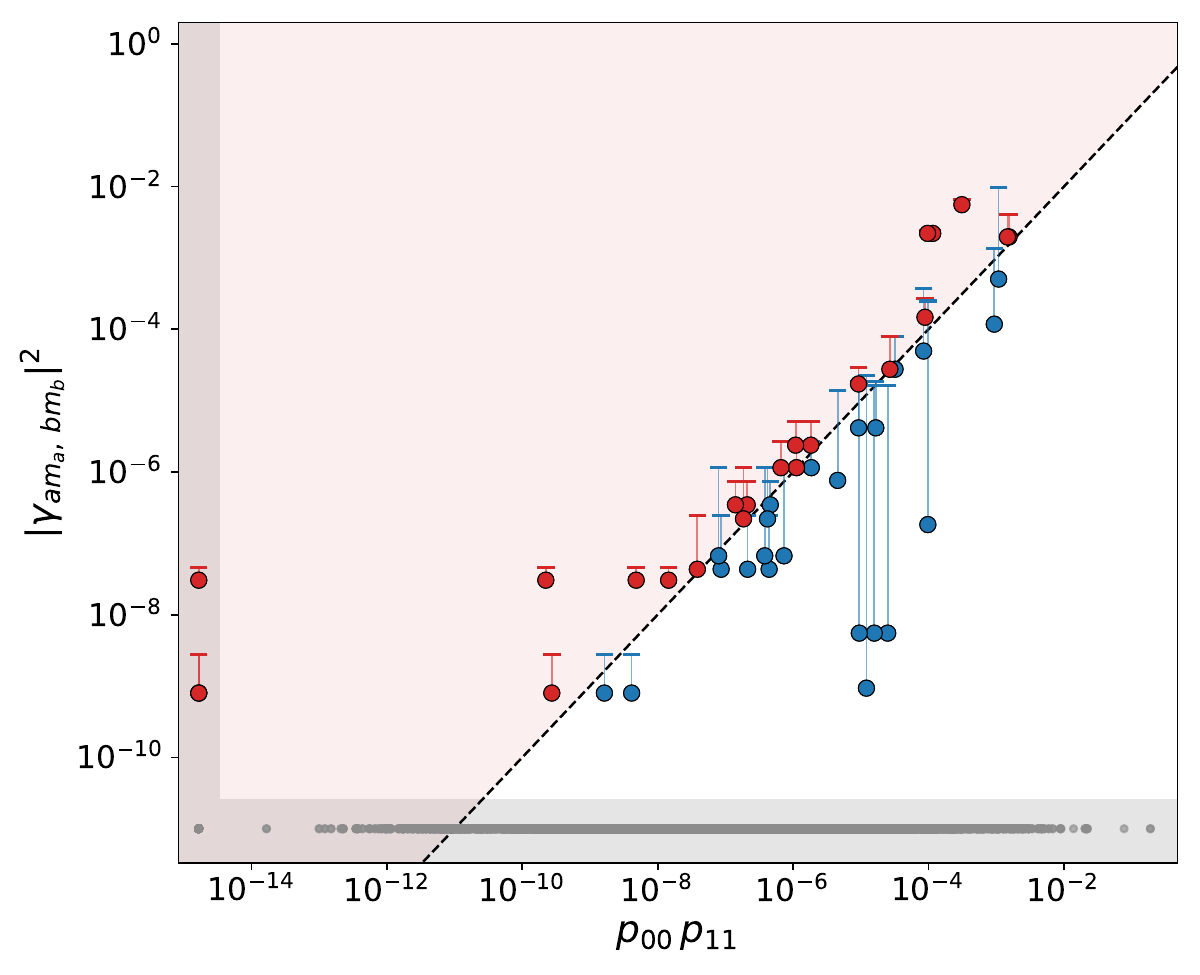} \\
    (b) \\
    \includegraphics[width=0.85\linewidth]{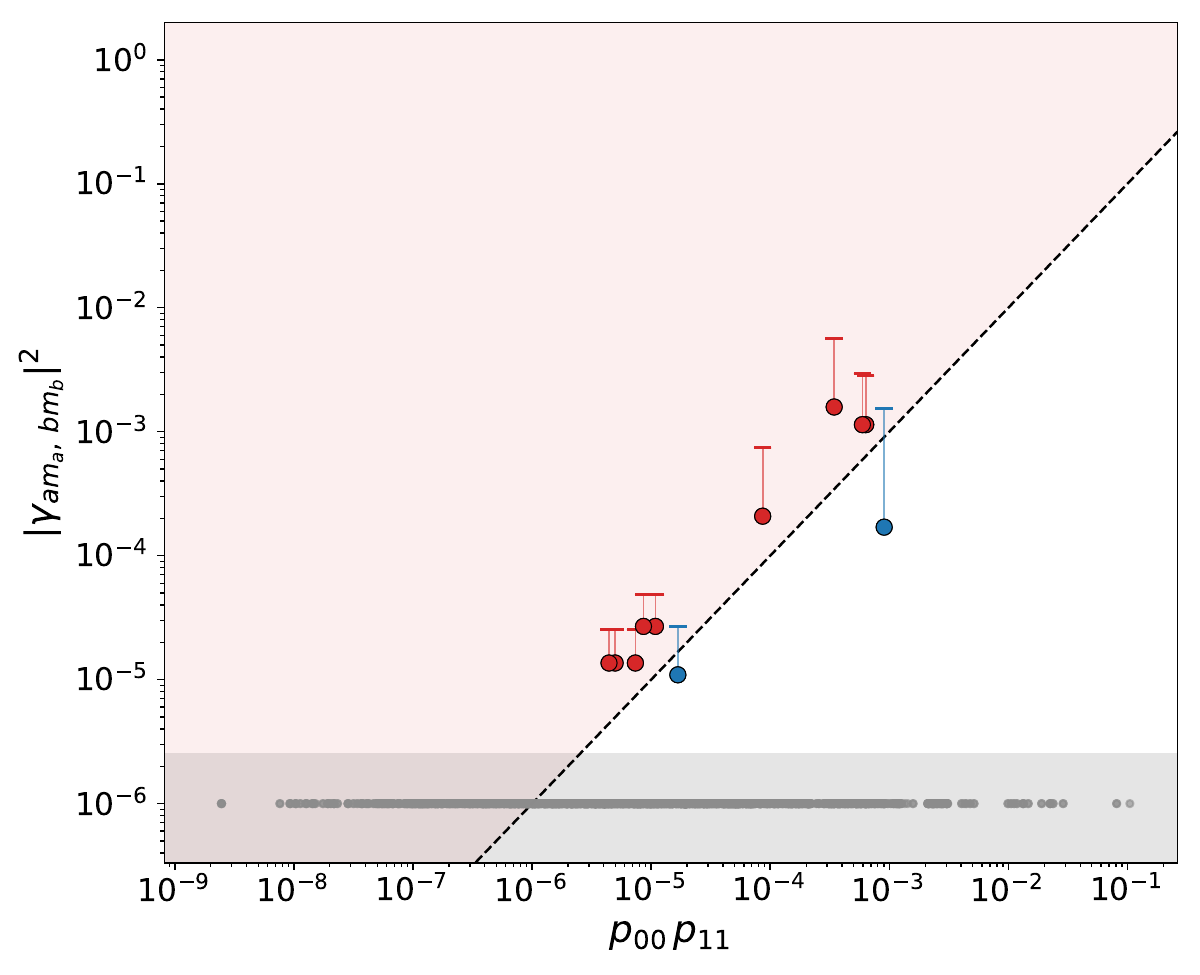}\\
    (c)\\

    \caption{
    Pair modes classified in three regions of correlations: entangled with $\mathcal{N}^{(\ama,\bmb)} > 0$ (red shade), classically correlated with $\gamma_{\ama,\bmb}=0$ (gray dots), and quantum but separable (blue dots), for (a) $^6$He (protons), (b) $^6$He (neutrons), and (c) $^{12}$C (neutrons), in the harmonic oscillator $j$-mode basis. The vertical lines show the last bound in Eq.~\eqref{eq:su3bounds}, giving the largest coherence that pair could have, given its own occupations $\gamma_{\ama}$ and $\gamma_{\bmb}$ and its pair term $\Gamma_{\ama,\bmb}$. 
    }
    \label{fig:entangledregion}
\end{figure}

Entangled modes have been illustrated for selected nuclei in Fig.~\ref{fig:entangledregion}, as one moves from $^6$He, with protons practically filling a shell and neutrons in the valence shell, to $^{12}$C. We note that, in the natural-orbital basis, $\gamma$ is diagonal and all two-mode negativities~\eqref{eq:su3Negativity} vanish. The entanglement identified here thus depend on the chosen mode basis. In the figure, the length of each line for the separable quantum regime shows how far the pair is from the maximum bound, given by the last inequality of Eq.~\eqref{eq:su3bounds}. A mode pair with its bar below the diagonal can never be entangled, no matter what its coherence is. A short line means the coherence nearly saturates the bound. This is the case, e.g., for two pair modes in $^6$He (protons) [panel (a)]: one of them, being close to the diagonal, is only barely separable, whereas the other one could be entangled given its occupations and pair term, but is not because its coherence is too small (we note all pair modes are classified for numerical precision of $10^{-11}$  used for $\Gamma$, which is set by the TBDME precision, and the figure shows the pair modes that cannot be classified at this precision by open circles). The neutron modes of $^6$He show a richer classification, given the two extra neutrons beyond the $^4$He (alpha) cluster. In ${}^{12}{\rm C}$, panel (c) shows fewer neutron pairs with resolved nonzero coherence, consistent with its description being closer to that of natural orbits, as discussed above (Fig.~\ref{fig:su2vonNeumannEntropy}), and still identifies entangled pairs above the diagonal.

\subsection{Information from modes about the register}
\label{sec:register}

Under Jordan--Wigner encoding~\cite{JordanWigner1928},
each mode $\ama$ is one qubit, so $\rho^{(\ama)}$ of Eq.~\eqref{eq:su3rho1} is a single-qubit reduced density matrix and Eq.~\eqref{eq:su3rho2} is a two-qubit one. The mode measures can be then used to inform:
\begin{enumerate}
\item \textbf{Freezing modes.} $\mathcal{S}_{\ama} \approx 0$ means the mode is essentially empty or essentially full, and it can be removed from the active space at quantified cost. Indeed, Fig.~\ref{fig:su2vonNeumannEntropy} shows that entropies are practically zero two shells above the valence shell (and two shells below, for heavier systems) as one goes from $^4$He to $^{20}$Ne.

\item \textbf{Ordering and connectivity.} DMRG orbital ordering and tensor-network topology are chosen by maximizing weighted mutual information~\cite{Legeza2003,Rissler2006,Legeza2015}; the same $I_{\ama,\bmb}$ guides qubit layout under hardware connectivity constraints. This is the quantity that carries the practical content, more than $\mathcal{S}_{\ama}$.

\item \textbf{Basis choice.} Spin-orbital modes, which naturally follow from the SU(3)-coupling scheme, and $j$-modes are related by a unitary transformation, which does not preserve the tensor-product decomposition $\bigotimes_{\ama} \mathbf{H}_{\ama}$ of Fock space that defines the mode subsystems, so their mode entropies differ for the same state. Which coupling scheme localizes correlation
better can be addressed in this framework.
\end{enumerate}

\section{Shells as subsystems}
\label{sec:subshell}

Let $\rho^{(a)}$ denote the reduced density matrix of all $\Oms_a$ modes of shell $a$
together, and $f_a$ its filling fraction:
\begin{equation}
    f_a = \frac{n_a}{\Oms_a} = \frac{\bar O^0_{aa}}{\sqrt{\Oms_a}} \in [0,1],
    \label{eq:su3fa}
\end{equation}
which is the scalar $(0\,0)(00)0$ part
of $\gamma_{aa}$ of Eq.~\eqref{eq:su3xJ0}, with $\sum_{\ama} f_a = \sum_a n_a = A$ from Eq.~\eqref{eq:su3na1}. One can then set an upper bound to the entropy from the filling alone. Subadditivity \cite{ArakiLieb1970}, and concavity of $\Hb$ applied to $f_a$ as the mean of the $\gamma_{\ama}$, give:
\begin{equation}
\mathcal{S}\big(\rho^{(a)}\big) \;\le\; \sum_{m_a} \mathcal{S}_{\ama} \;\le\; \Oms_a\, \Hb(f_a),
\label{eq:su3subshellbound}
\end{equation}
and the second bound holds in every single-particle basis of the subshell, including the natural one, because $\operatorname{Tr}\gamma_{aa} = n_a$ is basis independent. Hence $\Oms_a\Hb(f_a)$ is the largest entanglement shell $a$ can have with the rest of the nucleus given only its filling. It yields a freezing criterion from a single OBDME: if $f_a \approx 0$ or $1$, all $\Oms_a$ qubits of the shell may be frozen at quantified cost. There is no corresponding lower bound. 

For example, since a small value of $\Oms_a\Hb(f_a)$ certifies that the shell can be frozen, with a known cost, for $^{20}$Ne, the $0s$ shell with filling $f_0=0.99$ can be frozen (0.11 nats), along with shells $\eta_a \ge 5$ (0.17 nats together). However, the $1p$-shell of $^{20}$Ne cannot be frozen even at $f_1=0.96$, which for six modes for protons at $\Hb (0.96) \approx 0.17$ each add up to 1.0 nat. For comparison, the $0s$-filling $f_0=0.91$ for $^{6}$Be (protons) carries 0.60 nats. 

Furthermore, the subshell mutual information
$I(a{:}b) = \mathcal{S}(\rho^{(a)}) + \mathcal{S}(\rho^{(b)}) - \mathcal{S}(\rho^{(ab)})$ is the total
correlation as in Sec.~\ref{sec:modemeasures}, since $\rho^{(ab)}$ is mixed, but it supports three rigorous statements. First,
$I(a{:}b) = 0$ if and only if $\rho^{(ab)} = \rho^{(a)} \otimes \rho^{(b)}$, which excludes
entanglement. Second, $I$ bounds entanglement from above, that is, the product of the marginals is
separable, so the relative entropy of entanglement obeys
$E_R \le \mathcal{S}(\rho^{(ab)} \| \rho^{(a)} \otimes \rho^{(b)}) = I(a{:}b)$~\cite{Vedral1997}, and
the squashed entanglement obeys $E_{\rm sq} \le \tfrac12 I(a{:}b)$~\cite{Christandl2004}.
Third, a separable state is globally at least as disordered as locally,
$\mathcal{S}(\rho^{(ab)}) \ge \max\{\mathcal{S}(\rho^{(a)}), \mathcal{S}(\rho^{(b)})\}$~\cite{Cerf1997,Nielsen2001}, so:
\begin{equation}
\begin{gathered}
I(a{:}b) > \min\big\{\mathcal{S}(\rho^{(a)}),\, \mathcal{S}(\rho^{(b)})\big\}\\
\Longrightarrow\; \text{$a$ and $b$ are entangled},
\end{gathered}
\label{eq:su3Iwitness}
\end{equation}
or equivalently a negative conditional entropy $\mathcal{S}(a|b) < 0$ certifies entanglement. For
$0 < I(a{:}b) \le \min\{\mathcal{S}(\rho^{(a)}), \mathcal{S}(\rho^{(b)})\}$ the mutual information cannot decide.

The exact $I(a{:}b)$ needs many-body densities, but one- and two-body densities
bracket it. Mutual information cannot increase under partial trace, and
$I(a{:}b) \le 2\min\{\mathcal{S}(\rho^{(a)}), \mathcal{S}(\rho^{(b)})\}$~\cite{ArakiLieb1970}. Using \eqref{eq:su3subshellbound}, we have:
\begin{equation}
\max_{m_a m_b} I_{\ama,\bmb} \le I(a{:}b)
\le 2\min\big\{\Oms_a \Hb(f_a), \Oms_b \Hb(f_b)\big\}.
\label{eq:su3Ibracket}
\end{equation}
Sharper than either bound is a witness: if any mode pair $(\ama, \bmb)$ across the two
subshells is negativity-entangled by Eq.~\eqref{eq:su3neg}, the subshells are entangled, because
tracing out cannot create entanglement. The
converse fails, since subshells can be entangled collectively with no single mode pair
showing it. 

\begin{figure}
    \centering
    \includegraphics[width=\linewidth]{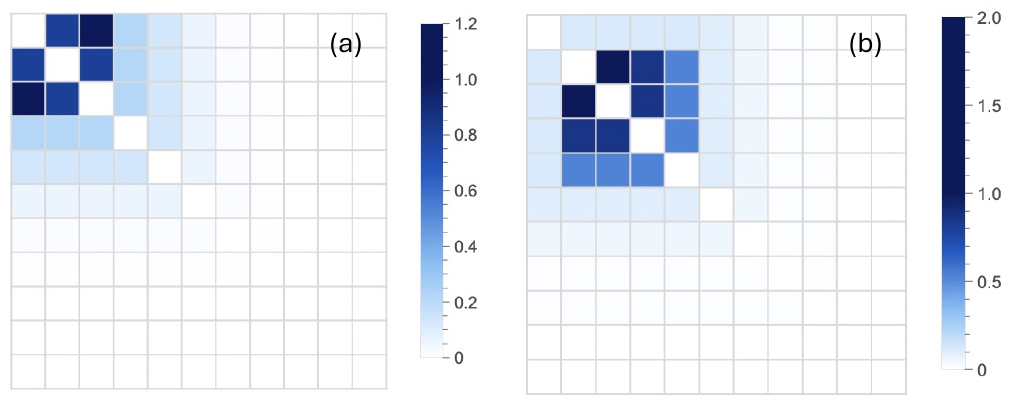}
    \caption{
    Upper-bound mutual information of two shells $I(a{:}b)$, for (a) $^6$He and (b) $^{20}$Ne for total of 11 shells, from proton one-body densities.
    }
    \label{fig:Iabmax}
\end{figure}

The upper-bound mutual information is illustrated for selected cases in Fig.~\ref{fig:Iabmax}, showing that for higher shells $I(a:b)$ becomes practically zero.  Zero values means not only there is no entanglement, but also no correlation of any kind, and no entanglement of either shell with the rest of the nucleus.
While, in practice, in many shells, the bound is small, not zero, it still bounds entanglement quantitatively, given the $E_R$ and $E_{\rm sq}$ above. For example, for $^4$He, the smallest case is $\eta=6$,  where $I(\eta_6:\cdot)  \le 0.090 $ nats for any other shell. That means $E_R \le 0.09$ and $E_{sq} \le 0.045$ between shell 6 and any other shell, which is a quantified freezing cost.

\subsection{Purity decomposition using one-body densities}

Only the total $\gamma_{\ama}$ carries an entropy. Each non-scalar $(\lambda \, \mu)(LS)J$ channel enters through a traceless $U^{(\lambda\,\mu)\kappa (LS)J0}_{aa}$, so its contribution sums to zero over the subshell and is necessarily negative on some modes. Hence, no per-channel reduced density matrix exists. However, the purity decomposes additively over channels. From Eqns.~\eqref{eq:su3ortho}and \eqref{eq:su3gamma}, it follows that:
\begin{equation}
\begin{aligned}
\operatorname{Tr}\gamma^2 &= \sum_{ab}\sum_{(\lambda\,\mu)\kappa LSJ} \big|\bar O^{(\lambda\,\mu)\kappa (LS)J0}_{ab}\big|^2,\\
\mathcal{S}_T \equiv \operatorname{Tr}\gamma- \operatorname{Tr}\gamma^2 &= A - \operatorname{Tr}\gamma^2 = \sum_i \tilde n_i\,(1-\tilde n_i)
\;\ge\; 0
\end{aligned}\;
\label{eq:su3purity}
\end{equation}
with $\tilde n_i$ the eigenvalues of the full one-body density matrix $\gamma$ (the natural-orbital occupation numbers, whose eigenvectors are the natural orbitals). For a single subshell the same argument gives $\operatorname{Tr}\gamma_{aa}^2 = \sum_{(\lambda\,\mu)\kappa LSJ}|\bar O^{(\lambda\,\mu)\kappa (LS)J0}_{aa}|^2$. $\mathcal{S}_T$ informs impurity, or equally, the degree of  mixing, and is half the mode-summed Tsallis entropy, given by linear entropy. For completeness, we will briefly outline next the relevant theory, along with the lower and upper bounds $\mathcal{S}_T$ can provide for $\mathcal{S}_1$.

\begin{figure*}
    \centering
    \includegraphics[width=0.85\linewidth]{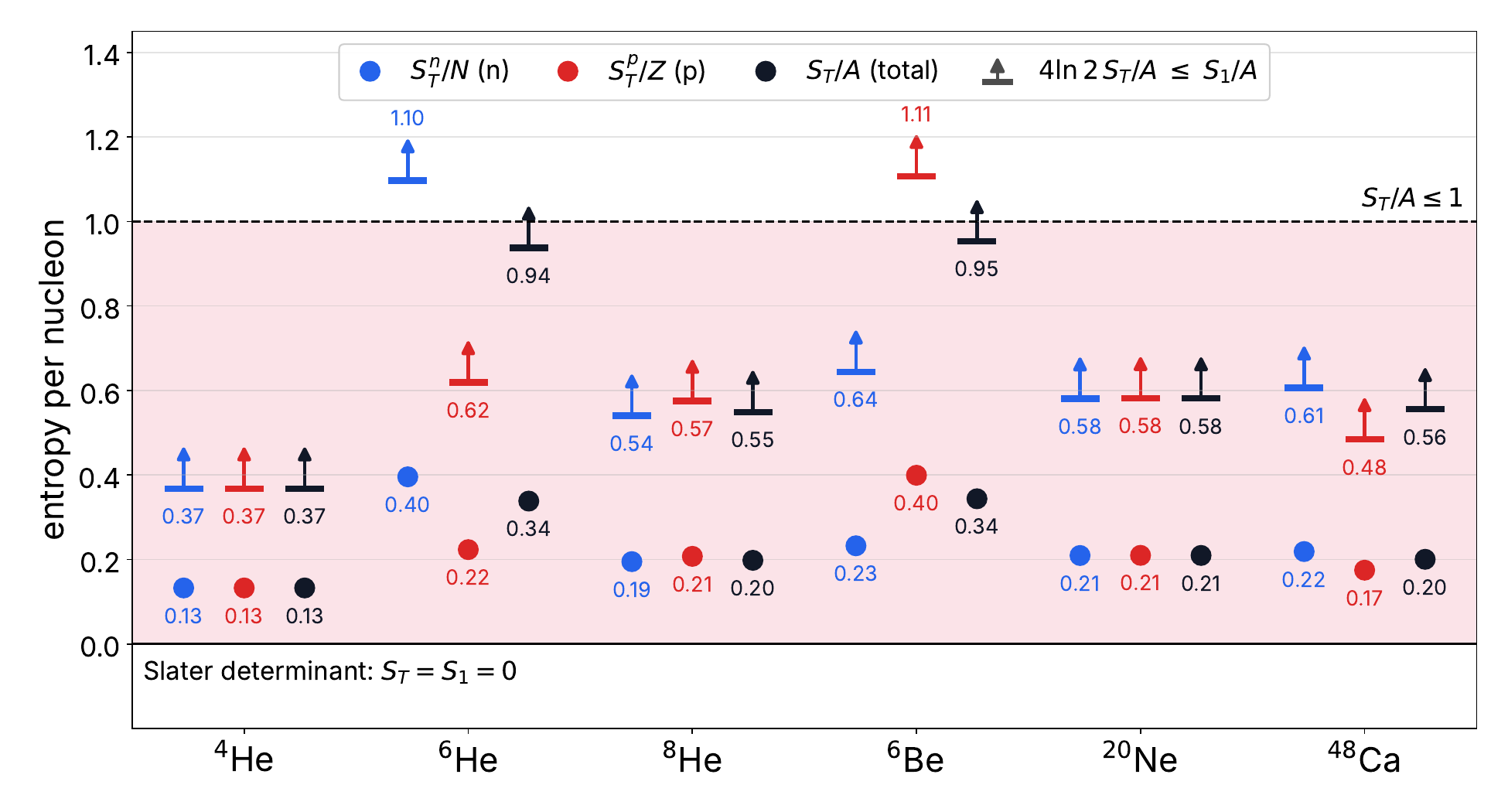}
    \caption{
    Entropy per nucleon, $S_T/N$ (neutrons, blue dots), $S_T/Z$ (protons, red dots), and $S_T/A$ (total, black dots) and the lower bound of $S_1/A$ (bars), for $^4$He through $^{48}$Ca, with model spaces detailed in Sec.~\ref{sec:SAbasis}. The shaded band is the range of $S_T/A$. 
    }
    \label{fig:entropy_per_nucleon}
\end{figure*}

Expanding the logarithm about $x = 1$, $-\ln x \approx 1 - x$, linearizes the von Neumann entropy:
\begin{equation}
\mathcal{S}(\rho) = -\operatorname{Tr}\rho\ln\rho \;\approx\; \operatorname{Tr}\rho(1-\rho)
= 1 - \operatorname{Tr}\rho^2 \;\equiv\; \mathcal{S}_L(\rho),
\label{eq:su3SL}
\end{equation}
where $\mathcal{S}_L$ is the linear entropy, which is the Tsallis entropy of order $q = 2$~\cite{Tsallis1988}
and is related to the second R\'enyi entropy by $\mathcal{S}_2 = -\ln(1 - \mathcal{S}_L)$. Like $\mathcal{S}(\rho)$ it vanishes exactly for a pure state and is largest for a maximally mixed one, but it is quadratic in $\rho$. This allows the decomposition over an orthonormal channel basis (Sec.~\ref{sec:modes}). For a single mode, $\mathcal{S}_L(\mathrm{diag}(1-h, h)) = 2h(1-h)$, the linearization of $\Hb(h)$.

Summing the Tsallis ($q = 2$) entropy, defined in Eq.~\eqref{eq:su3SL}, of each single-mode density matrix over the natural orbitals gives:
\begin{equation}
\begin{aligned}
\mathcal{S}_T &= \tfrac12 \sum_i \mathcal{S}_L\big(\rho^{(i)}\big)
= \tfrac12 \sum_i \Big[1 - (1-\tilde n_i)^2 - \tilde n_i^2\Big]\\
&= \sum_i \tilde n_i (1-\tilde n_i),
\end{aligned}
\label{eq:STsum}
\end{equation}
which is term by term the linear counterpart of $\mathcal{S}_1$ in Eq.~\eqref{eq:su3S1}. The leading $\tfrac12$ is a
normalization factor, and with it, each natural orbital contributes
$\tilde n_i(1-\tilde n_i)$, which has two direct meanings: it is the variance of the orbital's
occupation, $\operatorname{Var}(\hat N_i) = \tilde n_i(1-\tilde n_i)$, and it is
$\det\rho^{(i)}$, since $\mathcal{S}_L = 2\det\rho$ for any $2\times2$ density matrix. $\mathcal{S}_T$ is thus
the sum of the natural-orbital occupation variances. Its range is:
\begin{equation}
0 \;\le\; \mathcal{S}_T \;\le\; A\Big(1 - \frac{A}{\Om}\Big),
\label{eq:STrange}
\end{equation}
and reaches the maximum  when all $\tilde n_i = A/\Om$. The ratio
$\mathcal{S}_T/[A(1 - A/\Om)] \in [0,1]$, which approaches $\mathcal{S}_T/A$ in large model spaces, is a normalized distance from mean field that can be compared
across nuclei and model spaces (Fig.~\ref{fig:entropy_per_nucleon}).

As illustrated in Fig.~\ref{fig:entropy_per_nucleon}, for the nuclei considered, the per-nucleon $\mathcal{S}_T$ (for protons, neutrons, and total) sits between 0.13 and 0.40. That is far below its maximum of 1, but never close to the Slater-determinant value of 0. The distance from a determinant per nucleon depends mainly on how full the valence shell is (see the neutron entropy for the He isotopes), not on the mass number. Indeed, the entropies for $^8$He are very similar to those of $^{20}$Ne and $^{48}$Ca.

The use of the natural basis in Eq.~\eqref{eq:STsum} is essential, since, in any other basis, the mode sum $\sum_i \gamma_{ii}(1 - \gamma_{ii}) \ge \mathcal{S}_T$ (which follows from $\sum_i \gamma_{ii}^2 \le \operatorname{Tr}\gamma^2$) with equality only in the natural basis. $\mathcal{S}_T$ is therefore the basis-independent minimum of the mode-summed linear entropy, exactly as $\mathcal{S}_1$ is for the von Neumann entropy. This is important, since the two-mode entanglement discussed in Sec.~\ref{sec:modemeasures} is a property of the oscillator-basis register rather than of the nucleus alone: it measures the entanglement a circuit acting on oscillator modes must create, and is removed by a single-particle rotation to natural orbitals (a layer of Givens rotations~\cite{Kivlichan2018}). Correlation that persists in every single-particle basis is measured instead by $\mathcal{S}_T$ and also by $\mathcal{S}_1$.

Two inequalities for the binary entropy,
$4\ln 2\; h(1-h) \le \Hb(h) \le 2\ln 2\sqrt{h(1-h)}$, applied term by term connect it to the natural-orbital entropy of Eq.~\eqref{eq:su3S1}:
\begin{equation}
4\ln 2\; \mathcal{S}_T \;\le\; \mathcal{S}_1 \;\le\; 2\ln 2\,\sqrt{\Om\,\mathcal{S}_T},
\label{eq:STS1}
\end{equation}
where the upper bound uses the Cauchy--Schwarz inequality, and both are saturated when all $\tilde n_i = \tfrac12$, where $\mathcal{S}_T = \Om/4$ and $\mathcal{S}_1 = \Om\ln 2$. $\mathcal{S}_T$ thus bounds $\mathcal{S}_1$ from both sides without diagonalization of one-body densities, and the two vanish together. For identical fermions the linear entropy is often applied instead to the unit-trace density $\gamma/A$~\cite{Plastino2009}, the full one-body density matrix of Eq.~\eqref{eq:su3gammadef} normalized to unit trace---for a Slater determinant, it equals $1 - 1/A$ rather than zero, and $\mathcal{S}_T = A^2\big[\mathcal{S}_L(\gamma/A) - (1 - 1/A)\big]$.
 
Eq.~\eqref{eq:su3purity} provides entropy that can be calculated from the one-body densities only, and has the following important properties:
\begin{enumerate}
\item \textbf{It decomposes exactly over channels.} Each channel $(\lambda\,\mu) (LS)J$ and each shell pair
contributes one non-negative term with no cross terms. The von Neumann entropy has no such decomposition, because $-h\ln h$ is not quadratic in $\gamma$ while $\operatorname{Tr} \gamma^2$ is.

\item \textbf{It is basis independent.} $\operatorname{Tr}\gamma^2$ is invariant under any unitary rotation of the single-particle basis, whereas $\mathcal{S}_{\ama}$ changes between
$j$-level, HF, and natural bases. It therefore supplies a fixed reference against which
mode entropies computed in different bases can be compared.

\item \textbf{It measures correlation directly.} $\mathcal{S}_T$ vanishes if and only if $\gamma$ is idempotent ($\gamma = \gamma^2$), that is, if and only if the state is a single Slater determinant~\cite{Coleman1963}. It is the Tsallis counterpart of $\mathcal{S}_1$ in Eq.~\eqref{eq:su3S1} and bounds it from both sides according to Eq.~\eqref{eq:STS1} without diagonalization of one-body densities.

\item \textbf{It informs state preparation through one-body densities.} A Slater determinant can be prepared with a circuit of Givens rotations of linear depth~\cite{Kivlichan2018}, so $\mathcal{S}_T$ measures how far the target is from a state the hardware can prepare cheaply, and its channel decomposition says which multipoles carry that distance. Any correlation beyond the Slater determinant is what the variational part of the circuit must build.
\end{enumerate}

\section{Symmetry channel measures: informing the circuit}
\label{sec:channels}

\subsection{Pair numbers and the normalized Gram matrix} 

The channel pair-number operator:
\begin{equation}
\hat N^{(\lambda\,\mu)S\alpha }_{\mathrm P,ab} = \sum_M P^{(\lambda\,\mu)S\alpha M\dagger}_{ab}\, P^{(\lambda\,\mu)S\alpha M}_{ab}
\label{eq:su3NP}
\end{equation}
counts the pairs of configuration $ab$ in channel $(\lambda\,\mu)S\alpha$. Its expectation value gives the pair occupation of the configuration ($a \le b$) and channel, and is defined as:
\begin{equation}
n^{(\lambda\,\mu)S\alpha}_{ab} = \avg{\hat N^{(\lambda\,\mu)S\alpha }_{\mathrm P,ab}}
= \sum_M \bar O^{(\lambda\,\mu)S\alpha M,\,(\lambda\,\mu)S\alpha M}_{ab,\,ab},
\label{eq:su3nab}
\end{equation}
similarly to the shell occupation $n_a$, with a trace of:
\begin{equation}
    \sum_{a \le b}\, \sum_{(\lambda\,\mu)S\alpha} n^{(\lambda\,\mu)S\alpha}_{ab} = \frac{A(A-1)}{2}.
\label{eq:su3Npair}
\end{equation}
In Secs.~\ref{sec:channels}--\ref{sec:layout}, the pair densities are analyzed separately for $pp$ and $nn$ pairs. In the pair-density formulas, $A=Z$ or $N$, and $\gamma$ and $\mathcal{S}_T$ refer to the same species. The channel counts and operator pools below refer to like-particle pair excitations only.

More generally, the elements of $\bar O^{(\lambda\,\mu)S\alpha M,(\lambda'\, \mu')\kappa'(L'S')J'M}_{ab,cd}$  [Eq.~\eqref{eq:su3Obar2}] are the inner
products of the vectors $P^{(\lambda\,\mu)S\alpha M}_{ab}\ket{\Psi^{J_0M_0}}$, and they form the Gram matrix of these vectors, which is
Hermitian, positive semidefinite by construction, and diagonal in $M$. Normalized to unit trace, it is the normalized pair Gram matrix:
\begin{equation}
\begin{gathered}
\Gb^{(\lambda\,\mu)S\alpha M,\,(\lambda'\, \mu')S'\alpha'M}_{ab,\,a'b'}
= \frac{\bar O^{(\lambda\,\mu)S\alpha M,\,(\lambda'\, \mu')S'\alpha' M}_{ab,\,a'b'}}{A(A-1)/2},\\[3pt]
\sum_{a \le b}\, \sum_{(\lambda\,\mu)S\alpha M} \Gb^{(\lambda\,\mu)S\alpha M,\,(\lambda\, \mu)S\alpha M}_{ab,\,ab} = 1 .
\end{gathered}
\label{eq:su3gram}
\end{equation}
The normalized pair occupations are its $M$-summed diagonal:
\begin{equation}
\begin{gathered}
p^{(\lambda\,\mu)S\alpha}_{ab} = \frac{n^{(\lambda\,\mu)S\alpha}_{ab}}{A(A-1)/2}
= \sum_M \Gb^{(\lambda\,\mu)S\alpha M,\,(\lambda\,\mu)S\alpha M}_{ab,\,ab},\\
p^{(\lambda\,\mu)} = \sum_{a \le b}\, \sum_{S\alpha} p^{(\lambda\,\mu)S\alpha}_{ab}.
\end{gathered}
\label{eq:su3ppair}
\end{equation}
which are the probability that a pair drawn uniformly from all $A(A-1)/2$ pairs is in configuration $ab$
and channel $(\lambda\,\mu)S\alpha$.

\subsection{The channel space for two shells}

For two shells $a, b$ and one channel $(\lambda\,\mu)S\alpha M$, with a configuration set:
\begin{equation}
\mathcal{C}_{ab} = \{aa,\ ab,\ bb\},
\label{eq:su3Cset}
\end{equation}
the weight of the channel in these two shells is given through the eigenvalues 
$g^{(\lambda\,\mu)S\alpha M}_{i}$, $i = 1,2,3$, of the $3\times3$ block of $\Gb$ over $\mathcal{C}_{ab}$:
\begin{equation}
\begin{aligned}
W^{(\lambda\,\mu)S\alpha M}_{\mathcal{C}_{ab}} &= \sum_i g^{(\lambda\,\mu)S\alpha M}_{i}
= \operatorname{Tr}_{\mathcal{C}_{ab}} \Gb^{(\lambda\,\mu)S\alpha M,\,(\lambda\,\mu)S\alpha M}\\
&= \sum_{a'b' \in \mathcal{C}_{ab}} \Gb^{(\lambda\,\mu)S\alpha M,\,(\lambda\,\mu)S\alpha M}_{a'b',\,a'b'}.
\end{aligned}
\label{eq:su3WCtrace}
\end{equation}
The subscript $\mathcal{C}_{ab}$ refers to the whole set, which is different than a subscript $ab$ denoting a single configuration ($a\le b$), as in $p_{ab}$ in Eq.~\eqref{eq:su3ppair}, and $a'b'\in\mathcal C_{ab}$ runs over the members of the set. The channel weights and their sum over $M$, given by:
\begin{equation}
W^{(\lambda\,\mu)S\alpha}_{\mathcal{C}_{ab}} = \sum_M W^{(\lambda\,\mu)S\alpha M}_{\mathcal{C}_{ab}}
= p^{(\lambda\,\mu)S\alpha}_{aa} + p^{(\lambda\,\mu)S\alpha}_{ab} + p^{(\lambda\,\mu)S\alpha}_{bb},
\label{eq:su3wC}
\end{equation}
are the probability that a pair drawn from all $A(A-1)/2$ pairs lies in channel $(\lambda\,\mu)S\alpha M$, or $(\lambda\,\mu)S\alpha$, and in one of the configurations of $\mathcal{C}_{ab}$. $W^{(\lambda\,\mu)S\alpha}$ is a trace and hence unchanged by the coherences $\Gb_{aa,ab}, \Gb_{aa,bb}, \Gb_{ab,bb}$ between configurations. We note that $\sum_{a\le b} W_{\mathcal{C}_{ab}}$ is not one.

Adjoining a channel-empty state $\ket{\varnothing}$, which carries the complementary weight $1 - W^{c}_{\mathcal{C}_{ab}}$ of all pair strength outside both channel $c$ and configuration $\mathcal{C}_{ab}$, gives a four-dimensional space with ordered basis:
\[
\big\{ \ket{\varnothing},\ \ket{(aa)},\ \ket{(ab)},\ \ket{(bb)} \big\} .
\]
Writing $\Gb_{ab,a'b'}$, with $ab,a'b'\in\mathcal C_{ab}$, for the diagonal-channel elements $\Gb^{(\lambda\,\mu)S\alpha M,(\lambda\,\mu)S\alpha M}_{ab,a'b'}$, the matrix in this basis is:
\begin{equation}
\rho^{(\lambda\,\mu)S\alpha M}_{ab} =
\begin{pmatrix}
1 - W^{(\lambda\,\mu)S\alpha M}_{\mathcal{C}_{ab}} & 0 & 0 & 0\\
0 & \Gb_{aa,aa} & \Gb_{aa,ab} & \Gb_{aa,bb}\\
0 & \Gb_{ab,aa} & \Gb_{ab,ab} & \Gb_{ab,bb}\\
0 & \Gb_{bb,aa} & \Gb_{bb,ab} & \Gb_{bb,bb}
\end{pmatrix},
\label{eq:su3chan4}
\end{equation}
with eigenvalues $\{1-W^{(\lambda\,\mu)S\alpha M}_{\mathcal{C}_{ab}},\ g_i\}$. The $\ket{\varnothing}$ row and column are off-diagonally zero because the configuration states differ from the vacuum state by two particles. The $g_i$ are positive and normalized, so they form a categorical distribution, and entropy can be calculated as $\mathcal{S}_{\Gb} = -\sum_{i} g_i \ln g_i$. Beyond the weight, the eigenvectors of Eq.~\eqref{eq:su3chan4} give the natural pair states, namely, a single dominant $g_i$ means the channel is carried by one coherent combination of $(aa), (ab), (bb)$.

\subsection{Channel weights and SU(3) profiles}
\label{sec:weights}

Using Eq.~\eqref{eq:su3wC}, the joint channel weight of $(\lambda\,\mu)$ and the set, $W^{(\lambda\,\mu)}_{\mathcal{C}_{ab}} = \sum_{S\alpha} W^{(\lambda\,\mu)S\alpha}_{\mathcal{C}_{ab}} \equiv p((\lambda\,\mu),\mathcal{C}_{ab})$, gives the conditional profile and its entropy:
\begin{equation}
\begin{gathered}
w^{(\lambda \, \mu)}_{\mathcal{C}_{ab}} = \frac{W^{(\lambda \, \mu)}_{\mathcal{C}_{ab}}}{W_{\mathcal{C}_{ab}}}
= \frac{\sum_{S\alpha} W^{(\lambda\,\mu)S\alpha}_{\mathcal{C}_{ab}}}
{\sum_{\lambda'\mu'S\alpha} W^{(\lambda'\,\mu')S\alpha}_{\mathcal{C}_{ab}}} \equiv p((\lambda\,\mu)\,|\,\mathcal{C}_{ab}),\\[3pt]
\Hb_{\mathcal{C}_{ab}} = -\sum_{\lambda\mu} w^{(\lambda \, \mu)}_{\mathcal{C}_{ab}} \ln w^{(\lambda \, \mu)}_{\mathcal{C}_{ab}},
\end{gathered}
\label{eq:su3WL}
\end{equation}
that is, given that a pair lies in $\mathcal{C}_{ab}$, $w^{(\lambda \, \mu)}_{\mathcal{C}_{ab}}$ is the probability that the SU(3) rank of the pair is $(\lambda \, \mu)$, normalized to one within each set: $\sum_{\lambda\mu}w^{(\lambda \, \mu)}_{\mathcal{C}_{ab}}=1$. The effective number of contributing SU(3) channels in the set can be then calculated as:
\begin{equation}
N_{\text{eff}}(\mathcal{C}_{ab}) = e^{\Hb_{\mathcal{C}_{ab}}}
= \exp\Big(-\sum_{\lambda\mu} w^{(\lambda \, \mu)}_{\mathcal{C}_{ab}} \ln w^{(\lambda \, \mu)}_{\mathcal{C}_{ab}}\Big).
\label{eq:su3NeffC}
\end{equation}
We note that these formulae can be generalized for the $L$ profile (or $S$ profile), e.g., $W^{L}_{\mathcal{C}_{ab}} = \sum_{\lambda\,\mu \kappa SJ} W^{(\lambda\,\mu)\kappa (LS)J}_{\mathcal{C}_{ab}} \equiv p(L,\mathcal{C}_{ab})$ with the corresponding definitions for $w^L_{\mathcal{C}_{ab}}$ and the effective number of contributing $L$ channels in the set $N_{\text{eff}}(\mathcal{C}_{ab})=e^{-\sum_{L} w^{L}_{\mathcal{C}_{ab}} \ln w^{L}_{\mathcal{C}_{ab}}}$.

\begin{figure}
    \centering
    \includegraphics[width=1\linewidth]{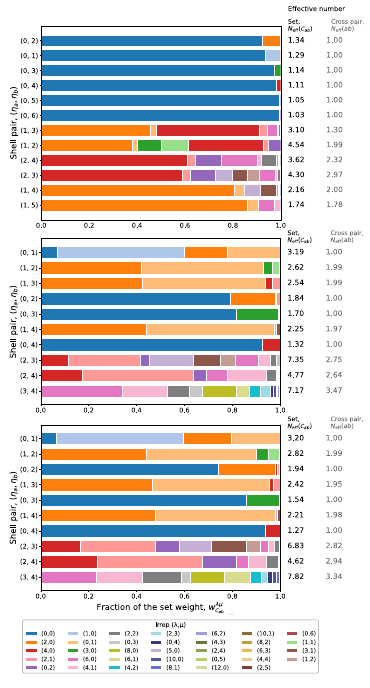}
    \caption{
    Profile $w^c_{\mathcal C_{ab}}$ of Eq.~\eqref{eq:su3WL} for the 10 shell pairs with the largest cross-configuration weight $p_{ab}$, with $N_{\rm eff}(\mathcal C_{ab})$ and the per-configuration $N_{\rm eff}(ab)$ for the neutrons in $^4$He (top), $^8$He (middle), and $^{12}$C (bottom), showing that the SU(3) sets are dominated by $(0,\eta)$ and $(1,\eta)$ configurations. This profile can be used to rank channels for the basis, register, and circuit initial state.
    }
    \label{fig:su3profile}
\end{figure}

Similarly, the global probability $p^{(\lambda \, \mu)}$, defined in Eq.~\eqref{eq:su3ppair}, yields the global effective number:
\begin{equation}
N_{\rm eff} = e^{\Hb(\{p^{(\lambda \, \mu)}\})},
\label{eq:su3Neff}
\end{equation}
which combines the spread of $(\lambda \, \mu)$ within configurations with the diversity between them, and can exceed every local count. Conversely, a weakly populated subshell pair can have a large local $N_{\rm eff}(\mathcal{C}_{ab})$ and barely register in $p^{(\lambda \, \mu)}$.  In addition, the per-configuration count is similarly defined as:
\begin{equation}
N_{\rm eff}(ab) = \exp\Big(-\sum_{\lambda \, \mu} w^{(\lambda \, \mu)}_{ab}\ln w^{(\lambda \, \mu)}_{ab}\Big),
\label{eq:su3Neffab}
\end{equation}
with the configurations $ab$ partitioning the pairs. The global count suits comparisons across nuclei, while the local ones provide decisions within a subshell pair, and both require the uncorrelated-reference subtraction detailed in Sec.~\ref{sec:ref}.

We illustrate the SU(3) profile for the neutron pairs in $^4$He, $^8$He, and $^{12}$C (Fig.~\ref{fig:su3profile}). The shell-pair weights: 
\begin{equation}
  p_{ab} = \sum_{(\lambda\,\mu)S\alpha} p^{(\lambda\,\mu)S\alpha}_{ab}  
  \label{eq:su3pab}
\end{equation}
are the same in any coupling scheme, as far as the weights are summed up to the shell subsystems (e.g., for $jj$, $p_{\eta_a \eta_b}=\sum_{\eta_a \le \eta_b} p_{\eta_a j_a ,\eta_b j_b}$). What differs is how each pair is resolved. In SU(3) every configuration $(0,\eta)$ belongs to a single irrep, $(\eta \,0)$, so a $0$-$\eta$ pair is one symmetry-coupled generator per spin channel, covering all orbits of shell $\eta$, whereas $jj$ spreads the same weight over orbit pairs and $J$. A different coupling scheme therefore suggests a different grouping of generators due to the basis used. We provide a comparison between the $jj$- and SU(3)-coupled basis in Sec.~\ref{sec:cmpWsu2su3}.

The SU(3) profile $w^c_{\mathcal C_{ab}}$ indicates where the pair strength is, including the uncorrelated part that a Gaussian circuit prepares cheaply. While $w^c_{\mathcal C_{ab}}$ and the effective number of channels ($N_{\rm eff}$) do not quantify the entanglement cost, they can be used to rank channels for the basis, register, and circuit initial state. Specifically, a coupling scheme or single-particle basis in which the weight concentrates in a few channels describes the like-particle pair distribution compactly, including its mean-field part. Indeed, the SU(3) sets are dominated by $(0,\eta)$ and $(1,\eta)$ configurations. Informing the register, the SU(3) profile determines configurations that hold pairs, thereby including their modes on the register, whereas configurations with negligible weight are candidates to drop. While occupancy decides this, correlated or not, this is the two-body counterpart of the one-body freezing tests. Finally, the circuit initial state is supposed to carry most of the weight cheaply, and a dominant channel, or a single dominant $g_i$~\eqref{eq:su3chan4} within it, informs the content of that state.

\subsection{Cumulants and correlation change: ranking operators for the entangling part of the circuit} 
\label{sec:ref}

The $w^{(\lambda \, \mu)}_{\mathcal{C}_{ab}}$ profile is not by itself evidence for correlation. For a single Slater determinant, Wick's theorem plus recoupling distributes
the antisymmetrized product of two occupied orbitals over SU(3) irreps (or $L$) through the Clebsch-Gordan geometry
alone, giving $N_{\text{eff}}(\mathcal{C}_{ab}) > 1$ from a state with no correlation at all. This is the
statistics-driven baseline that Ghirardi \emph{et al.}~\cite{Ghirardi2002} and Eckert
\emph{et al.}~\cite{Eckert2002} argue should be called a correlation rather than
entanglement. Hence, one can define a pair-occupation cumulant $\Delta n^{(\lambda \, \mu)S\alpha}_{ab} = n^{(\lambda \, \mu)S\alpha}_{ab} - n^{(\lambda \, \mu)S\alpha; 0}_{ab}$ for Eq.~\eqref{eq:su3nab}, as well as its normalized counterpart of Eq.~\eqref{eq:su3ppair}:
\begin{equation}
\Delta p^{(\lambda \, \mu)S\alpha}_{ab} = p^{(\lambda \, \mu)S\alpha}_{ab} - p^{(\lambda \, \mu)S\alpha; 0}_{ab},
\label{eq:su3dp}
\end{equation}
where the uncorrelated reference, denoted by ``0", is subtracted. This is exactly the two-body cumulant, the part of the pair occupation not explained by the one-body density, revealing correlations beyond a single-particle scenario. The uncorrelated reference is built by Wick contraction of the same many-body state's one-body density of
Eq.~\eqref{eq:su3gamma} with the same occupations, but no correlation. The uncorrelated reference is basis independent and it can be constructed using the $\ama$ basis. Its pair density matrix element is the antisymmetrized product of $\gamma$ with itself,
$\avg{\hat N_{\ama} \hat N_{\bmb}}_0 = \gamma_{\ama}\gamma_{\bmb} - |\gamma_{\ama,\bmb}|^2$. Summing over all pairs:
\begin{equation}
\sum_{\ama<\bmb} \avg{\hat N_{\ama} \hat N_{\bmb}}_0 = \tfrac12\big(A^2 - \operatorname{Tr}\gamma^2\big)
= \frac{A(A-1)}{2} 
+ \frac{\mathcal{S}_T}{2}.
\label{eq:su3cumtrace}
\end{equation}
This means that the cumulant $\Delta n$ carries total pair weight 
$-\mathcal{S}_T/2$,
that is, the reference overcounts pairs by half the
Tsallis entropy, defined in Eq.~\eqref{eq:su3purity}. 

Therefore, the informative quantity is:
\begin{equation}
\Delta w^{(\lambda \, \mu)}_{\mathcal{C}_{ab}} = w^{(\lambda \, \mu)}_{\mathcal{C}_{ab}} - w^{(\lambda \, \mu),\,0}_{\mathcal{C}_{ab}},
\label{eq:su3dWL}
\end{equation}
where unnormalized channel differences $\Delta W^c_{\mathcal{C}_{ab}}$ must be read with the 
$-\mathcal{S}_T/2$
offset
in mind, while the conditional $w^{(\lambda \, \mu)}_{\mathcal{C}_{ab}}$ absorbs it by normalization. 

\begin{figure*}
    \centering
    \includegraphics[width=1\linewidth]{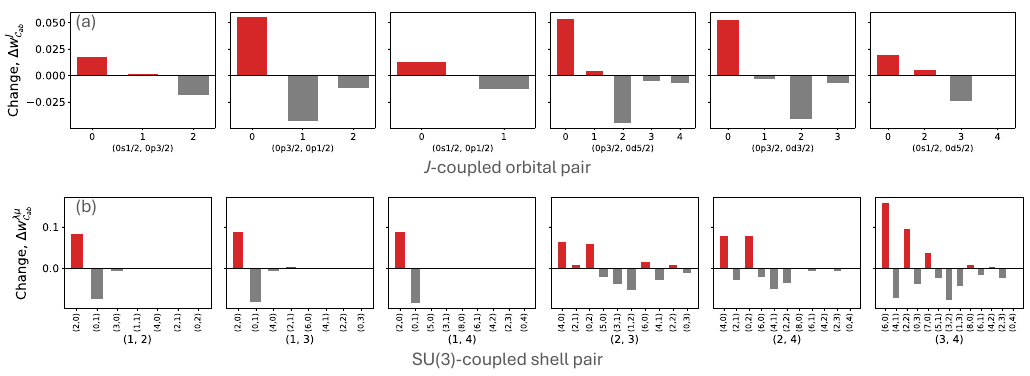}
    \caption{
$\Delta w^{c}_{\mathcal C_{ab}}$ vs. channels $c$ for the six leading  (a) pairs of orbits $(j_a,j_b)$ for the $jj$-coupling scheme ($c=J$), and (b) pairs of shells $(\eta_a,\eta_b)$ with both $\eta\ge1$ for the SU(3)-coupling scheme [$c=(\lambda\,\mu)$], for $^{12}$C  (neutrons). For SU(3), the correlation gains
sit mainly in $(\lambda \,0)$ with even $\lambda$, $S=0$, while irreps with $\mu>0$, along with
the $S=1$ channels lose weight. }
    \label{fig:W_deltaW}
\end{figure*}

For the SU(3) scheme, the gains and losses, given by $\Delta w$, are illustrated for $^{12}$C in Fig.~\ref{fig:W_deltaW}. While the pair weights themselves are large in many channels,
only a few channels have nonnegligible $\Delta w$. A global view follows from the
per-channel contributions $\mathcal{S}_T^c=-2\,\Delta n^c$ to the Tsallis entropy, that is,
by Eq.~\eqref{eq:su3cumtrace} they sum to $\mathcal{S}_T$ (or its numerical sum within the shells retained in the OBDME). A channel with
$\mathcal{S}_T^{c}>0$ loses pairs relative to the uncorrelated reference, and one with
$\mathcal{S}_T^{c}<0$ gains them. The leading irreps in $^{12}$C (neutrons) contribute to the total Tsallis entropy as follows:
$(0\,1)$: $+1.59$ (94.1\%); $(2\,0)$: $-0.41$ ($-24.1$\%); $(3\,0)$: $+0.20$
(11.9\%); $(4\,0)$: $+0.12$ (7.2\%); $(1\,1)$: $+0.09$ (5.5\%); $(1\,0)$: $+0.04$
(2.5\%); $(5\,0)$: $+0.03$ (1.8\%); $(3\,1)$: $+0.01$ (0.9\%), with a clear gain in correlations in $(2\,0)$ with $S=0$. The same pattern holds in $^{20}$Ne, where the gain sits in $(4\,0)$ with $S=0$.
In both nuclei, the correlation populates the pair irrep $(2\eta_v\,0)$ of the
valence shell $\eta_v$, which for like nucleons is the spatially symmetric $S=0$
combination with the largest deformation, at the expense of the $S=1$ channels and
the irreps with $\mu>0$.

\subsection{Information from symmetry-guided measures about the circuit}
\label{sec:circuit}

The symmetry-channel measures become very useful in the following cases:
\begin{enumerate}
    \item \textbf{The circuit content.} The channel weights split the circuit into the part that is
    cheap to prepare and the part that must be built. The set weight $W_{\mathcal{C}_{ab}}$ says which pairs of subsystems hold pairs at all, hence which modes the register must keep, and the profile $w^{c}_{\mathcal{C}_{ab}}$ says in which channels they sit, determining the channel content of the initial state, described compactly in a basis where few channels carry it. If one eigenvalue $g_i$ dominates a channel, the corresponding natural pair state is a good building block for that state (a pair condensate rather than a single determinant), which moves correlation out of the variational part of the circuit and into its initialization. The profile $\Delta w^{c}_{\mathcal{C}_{ab}}$ removes the part a determinant already supplies: channels with appreciable $|\Delta w|$, filled or emptied by the correlation, are candidates for pair excitations in $U(\theta)$. Both profiles select channels within a set but cannot rank sets.
    
    \item \textbf{Operator pool.} ADAPT-VQE~\cite{Grimsley2019} grows the ansatz from a pool of candidate excitations. Within a channel, the pair Gram matrix satisfies $|G_{ij}|^2\leq G_{ii}G_{jj}$, so diagonal occupations bound the pair-transfer density between configurations $i$ and $j$. The cumulant ranking identifies candidate channels, and ADAPT selects generators from the resulting pool by their energy gradients. Restricting the like-particle pool to the channels $c$ and configurations $ab$ with appreciable cumulant $|\Delta p^{c}_{ab}|$, for instance the smallest set that carries 99\% of $\sum|\Delta p|$, removes candidates before any gradient is measured, which is a large part of the measurement cost of ADAPT. The effective number $e^{\Hb(\{|\Delta p|\})}$ estimates the typical size of this pool, as a measure of concentration rather than a truncation rule (Sec.~\ref{sec:cmpWsu2su3}). The same channel set fixes the register graph (Sec.~\ref{sec:layout}) -- since adjacency matters for the gates applied, the pool and the layout are best chosen together.
    
    \item \textbf{Truncation with a quantified cost.} Because $\sum_{a\le b}\sum_{c} p^{c}_{ab} = 1$,
    the discarded share of the pair occupations, $D_p$ (Sec.~\ref{app:su3trunc}), estimates the fraction left out by a channel truncation of unitary coupled cluster, or of any pair-excitation
    ansatz. For the entangling operators, and for the pool cut above, the discarded cumulant $D_\Delta$, discussed in Sec.~\ref{app:su3trunc}, is the sharper measure. Dropping a channel $c$ removes all $\dim_c$ of its $m_c$ components, and every configuration in it, at once.
    
    \item \textbf{Symmetry preservation.} Each coupled pair operator $P^{(\lambda\mu)S\alpha M\dagger}_{ab}$ carries good $(\lambda\mu)S\alpha$, so an ansatz built from retained channels stays in the symmetry sector of the target without projection or penalty terms. Symmetry-preserving circuits are shallower and have fewer parameters~\cite{Gard2020}, and symmetry-adapted bases are used for nuclear quantum simulation for similar reasons~\cite{mumma2024efficaciousband,In2026symmetry}.
\end{enumerate}

None of these measures is circuit cost: a heavily weighted channel may be cheap to implement and a
lightly weighted one expensive, depending on the gate set and connectivity. 
The parameter counts in Sec.~\ref{app:su3trunc} assume one  generator per retained channel, a fixed combination of its $\dim_c$ operator components, applied once with a single  variational parameter.
The measures determine what the circuit must contain, in its initial state and in its entangling part, not its depth.

\begin{figure}
    \centering
    \includegraphics[width=1\linewidth]{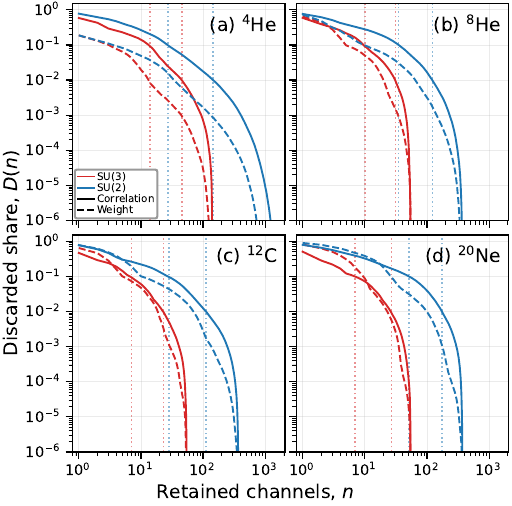}
    \caption{Discarded share $D_p(n)$ (dashed, labeled as ``Weight") and $D_\Delta(n)$ (solid, labeled as ``Correlation") vs. the number of retained channels for the SU(3) (red) and $jj$ (blue) coupling scheme. The number of retained channels at the 90\% (99\%) level corresponding to $D(n)=0.1$ ($D(n)=0.01$) are marked by vertical dotted lines, and they are smaller in the SU(3) scheme.  
    }
    \label{fig:Dshare}
\end{figure}

\subsection{Truncation with a quantified cost}
\label{app:su3trunc}

The effective number of channels $N_{\rm eff}$ measures how concentrated the distribution over channels is, but it is not a truncation rule. The number of channels to keep follows from the discarded share. Let the channels $k=(ab,(\lambda\mu)S)$ [or $(ab,J)$ in $jj$] be sorted by decreasing score and the first $n$ kept. Then:
\begin{equation}
\begin{gathered}
D_p(n)=1-\frac{\sum_{i\le n}p_{k_i}}{\sum_kp_k},\\
D_\Delta(n)=1-\frac{\sum_{i\le n}|\Delta p_{k_i}|}{\sum_k|\Delta p_k|},
\end{gathered}
\label{eq:Dtrunc}
\end{equation}
ranked by $p_k$ and by $|\Delta p_k|$ respectively (Fig.~\ref{fig:Dshare}). The count for a share $1-\epsilon$ kept is the smallest $n$ with $D(n)<\epsilon$: $\epsilon=10^{-1}$ for 90\%, $10^{-2}$ for 99\%, and so on (see vertical dotted lines in Fig.~\ref{fig:Dshare}). $D_\Delta$ counts gains and losses by magnitude. A channel that is empty in the state but populated in the reference enters $D_\Delta$ but not $D_p$. The nominal tensor-component (or operator) count of the retained channels is $\sum_{i\le n}\dim_c^{(k_i)}$, with $\dim_c$ for a channel $c$: $\dim(\lambda,\mu)\Pi^2_S$ for SU(3) irreps,
$2J+1$ for a $(\lambda\mu)\kappa LSJ$ multiplet and for $J$ irreps. 
\begin{figure*}
    \centering
    \includegraphics[width=0.85\linewidth]{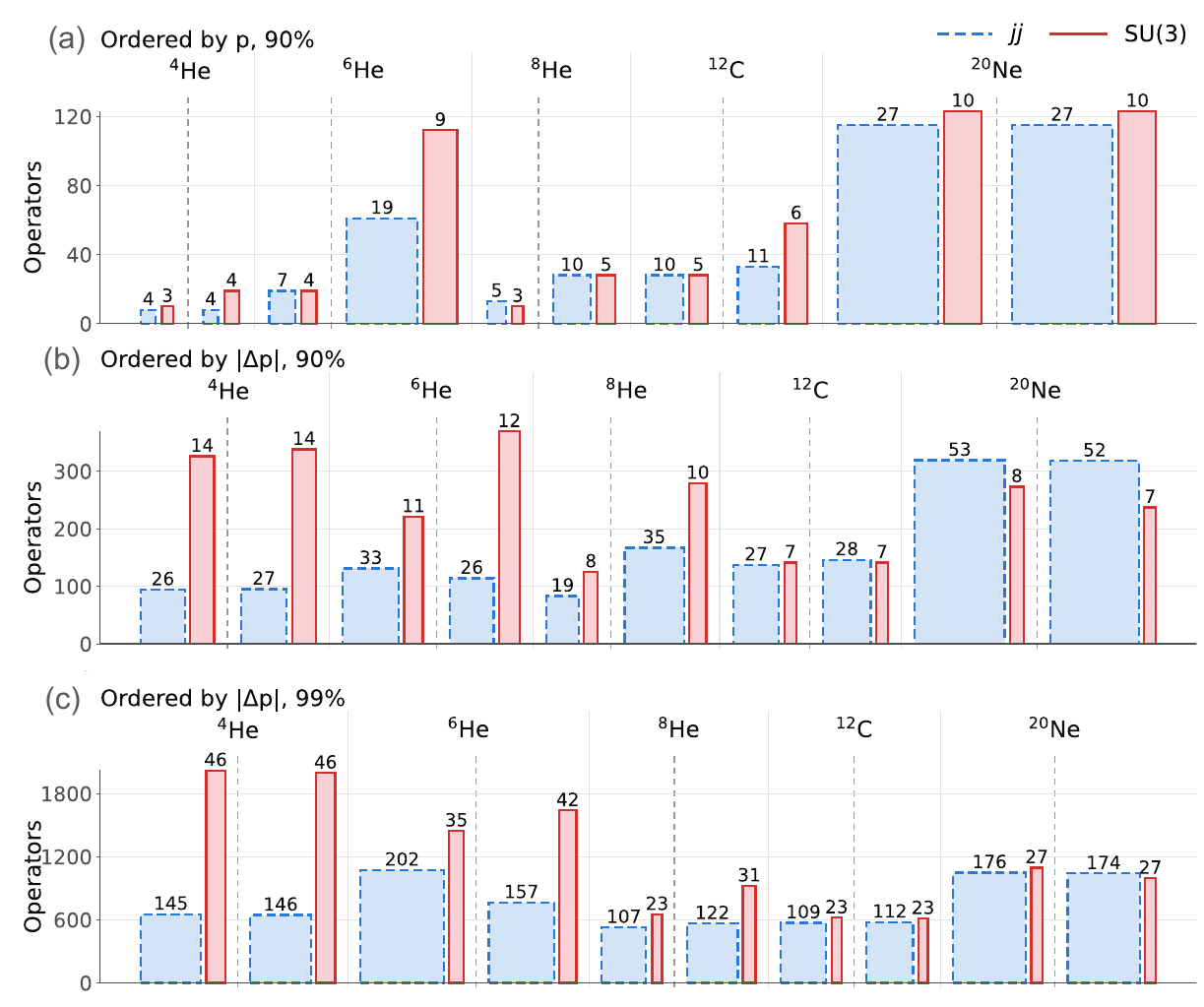}
    \caption{Channels and operators needed to achieve 90\% of (a) the total weight (determined by $D_p$) and (b) the total correlation (determined by $D_\Delta$), compared in (c) for 99\%.  SU(3)-operator counts are upper bounds, since components of an irrep may be zero due to finer selection rules. Results are shown across nuclei, for protons (neutrons) to the left (right) of each dashed vertical line. With fewer channels, the SU(3) scheme largely reduces the number of parameters and of candidate gradients per adaptive step at a comparable gate count, showing a clear advantage for the SU(3)-scheme compared to the $jj$-scheme for heavier systems. 
    }
    \label{fig:truncation90}
\end{figure*}

We compare the number of channels retained at 90\% and 99\% of the total weight ($D_p$) and of $\sum|\Delta p|$ ($D_\Delta$), and the number of like-particle pair-operator components they contain, across nuclei in the SU(3) and $jj$-coupled schemes (Fig.~\ref{fig:truncation90}). In every case shown, SU(3) retains fewer channels than $jj$, by factors of about 2 to 7. The operator counts are comparable in the two schemes. Since the SU(3) operator counts are upper bounds, these comparisons are conservative for SU(3), and in some cases SU(3) also retains fewer components.For example, compared to $jj$, ${}^{20}\mathrm{Ne}$ at 90\% of the cumulant strength has fewer SU(3) components (about 270 vs. 320 for protons and 240 vs. 320 for neutrons). The two counts measure different costs. In a symmetry-adapted ansatz, all components of a channel share one amplitude, so the number of channels sets the number of variational parameters and, in an adaptive algorithm, the number of candidate gradients screened per step. The number of operator components sets the number of pair-excitation gates, that is, the circuit size and an upper bound on its depth before parallelization, and the number of terms entering each screened gradient.  For the nuclei studied here, SU(3) therefore reduces the number of parameters and of candidate gradients per adaptive step at a comparable gate count, showing a robust advantage of SU(3).

We note that $D_p$ and $D_\Delta$ guide different choices in constructing the circuit.
$D_p$ ranks channels by where the pair strength sits, including the uncorrelated part that a Gaussian layer prepares cheaply. The rapid decreases of $D_p(n)$ shows that a small set of channels carries most of the pair weight. $D_\Delta$ ranks channels by the pair strength gained or lost relative to the uncorrelated reference with the same $\gamma$. Channels with large changes are candidates for the pair excitations that build correlations beyond the reference. For example, the leading cumulant channel in $^4$He, 0--1 in $(1,0)$ with $S=1$, alone carries 42\% of $\sum|\Delta p|$ and is not among the dominant channels by weight. The two rankings thus distinguish where the like-particle pair content resides from where it changes most. Both separate the SU(3) and $jj$ schemes for heavier nuclei studied here. We note that neither curve bounds the fidelity. A discarded share of $\epsilon$ in $|\Delta p|$ is a resource proxy, not an infidelity of $\epsilon$.

\section{Pair-channel entropies for SU(3)-channels and comparisons of SU(3) and $jj$ coupling schemes} 
\label{sec:cmpWsu2su3}

For two shells $\eta_a$ and $\eta_b$, the pair irreps are:
\begin{equation}
(\lambda\quad\mu)=(\eta_a+\eta_b-2k \quad k),\qquad k=0,\dots,\min(\eta_a,\eta_b),
\label{eq:su3pairs}
\end{equation}
and hence we have the constraint:
\begin{equation}
    \lambda+2\mu=\eta_a+\eta_b.
\end{equation}
According to Eq.~\eqref{eq:su3pairs}, the irreps of $aa$, $ab$ and $bb$ never coincide when $a\ne b$, since $\lambda+2\mu$ equals $2\eta_a$, $\eta_a+\eta_b$ and $2\eta_b$. Therefore, each $(\lambda\mu)$ of $\mathcal C_{ab}$ belongs to exactly one configuration, and the chain rule splits $\Hb_{\mathcal C_{ab}}$ exactly:
\begin{equation}
\begin{gathered}
\Hb_{\mathcal C_{ab}}=\Hb(\{q_{a'b'}\})+\sum_{a'b'\in\mathcal C_{ab}}q_{a'b'}\,\Hb(\{w^{(\lambda\,\mu)}_{a'b'}\}),\\
q_{a'b'}=\frac{p_{a'b'}}{W_{\mathcal C_{ab}}}\equiv p(a'b'\,|\,\mathcal C_{ab}) .
\end{gathered}
\label{eq:su3Hsplit}
\end{equation}
$N_{\rm eff}(\mathcal C_{ab})$ thus mixes how the weight is shared among $aa$, $ab$ and $bb$ with the spread over irreps inside each configuration. $N_{\rm eff}(ab)$ isolates the second, for the cross configuration alone. The same formulae hold when $(\lambda\mu)S$ is replaced by $(\lambda\mu)LSJ$ (sum over $\kappa M$ only) or by
$(\lambda\mu)\kappa LSJ$ (sum over $M$ only).

Further, we can explore global effective numbers. With $p_{ab}$ defined in Eq.~\eqref{eq:su3pab}, the fraction of pairs in configuration $(ab)$, $p^{(\lambda \, \mu)}_{ab} = \sum_{SJ} p^{(\lambda\,\mu)S\alpha}_{ab} \equiv p((\lambda\,\mu), ab)$, the joint weight of $(\lambda\,\mu)$ and the configuration, and
\begin{equation}
w^{(\lambda \, \mu)}_{ab} = \frac{p^{(\lambda \, \mu)}_{ab}}{p_{ab}} \equiv p((\lambda\,\mu)\,|\,ab)
\label{eq:su3wLab}
\end{equation}
the distribution of SU(3) rank within it, the following holds:
\begin{equation}
p^{(\lambda \, \mu)} = \sum_{a \le b} p^{(\lambda \, \mu)}_{ab} = \sum_{a \le b} p_{ab}\, w^{(\lambda \, \mu)}_{ab}.
\label{eq:su3pLW}
\end{equation}
While $w^{(\lambda \, \mu)}_{\mathcal{C}_{ab}}$ of Eq.~\eqref{eq:su3WL} is local and conditional (every subshell pair counts equally, however weakly populated), $p^{(\lambda \, \mu)}$ is global and weighted by the actual share of pairs. Treating the configuration and $(\lambda \, \mu)$ of a randomly drawn pair as two random variables, the chain rule gives:
\begin{equation}
\Hb\big(\{p^{(\lambda \, \mu)}\}\big) = \sum_{a \le b} p_{ab}\, \Hb\big(\{w^{(\lambda \, \mu)}_{ab}\}\big) + I((\lambda \, \mu);\,ab),
\label{eq:su3HLcd}
\end{equation}
with $I((\lambda \, \mu);\,ab) \ge 0$ the classical mutual information between SU(3) rank and configuration. In SU(3) the irrep and the configuration nearly determine each other ($\lambda+2\mu=\eta_a+\eta_b$), so $I(\lambda\mu;ab)$ nearly equals the entropy of the configurations, and the global effective number, $N_{\rm eff}=e^{\Hb(\{p^{\lambda\mu}\})}$ of Eq.~\eqref{eq:su3Neff}, counts mainly distinct configurations, not irreps within them. In $jj$ the roles are reversed: each orbit pair allows several $J$ and each $J$ occurs in many orbit pairs, so $I(J;ab)$ is small and $N_{\rm eff}$ counts mainly the spread of $J$ within configurations. A small $N_{\rm eff}$ identifies more concentrated channel structure to guide ansatz construction.

\begin{figure}
    \centering
    \includegraphics[width=0.85\linewidth]{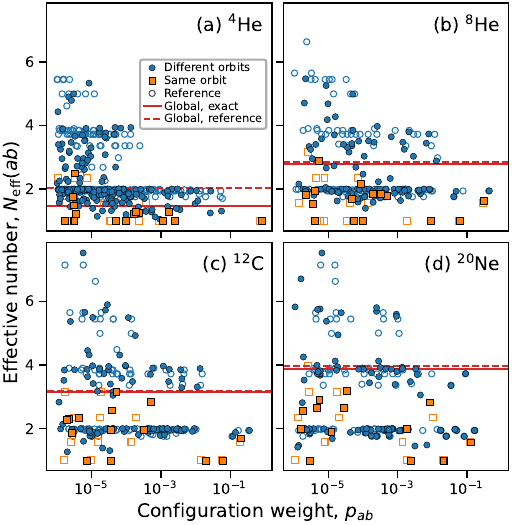}
    \caption{
    Effective number of pair angular momenta $J$ per configuration, $N_{\rm eff}(ab)$, vs. the configuration weight $p_{ab}$, for the neutron densities of (a) $^4$He, (b) $^8$He, (c) $^{12}$C and (d) $^{20}$Ne. Each point is one configuration $ab$, 
    a pair of $jj$ orbits (circles for different orbits; squares for the same orbit). Filled symbols: correlated state; open symbols: uncorrelated reference. Horizontal lines: global effective number $N_{\rm eff}$ over all configurations, for the state (solid) and the reference (dashed). Configurations with $p_{ab}<10^{-6}$ are not shown.
    }
    \label{fig:wvspjj}
\end{figure}

\begin{figure}
    \centering
    \includegraphics[width=0.85\linewidth]{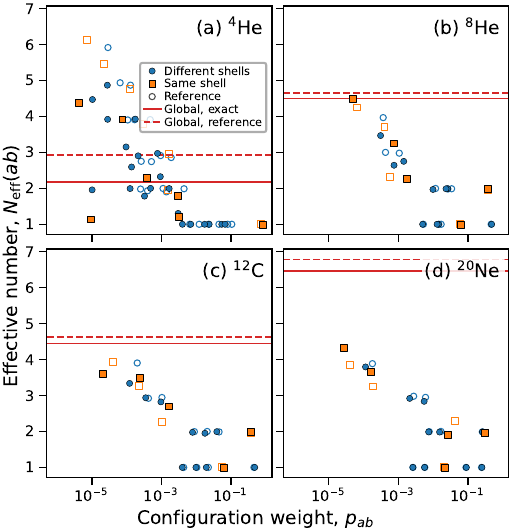}
    \caption{
    Same as in Fig.~\ref{fig:wvspjj}, but for the SU(3) coupling scheme. Effective number of SU(3) irreps per configuration, $N_{\rm eff}(ab)$, vs. the configuration weight $p_{ab}$, for the neutron densities of (a) $^4$He, (b) $^8$He, (c) $^{12}$C and (d) $^{20}$Ne. Each point is one configuration $ab$, a pair of oscillator shells $(\eta_a,\eta_b)$: circles for different shells, squares for the same shell. Filled symbols: correlated state; open symbols: uncorrelated reference. Horizontal lines: global effective number $N_{\rm eff}$ over all configurations, for the state (solid) and the reference (dashed). Configurations with $p_{ab}<10^{-6}$ are not shown.
}
    \label{fig:wvspSU3}
\end{figure}

Effective numbers are a tolerance-free measure of concentration. Examining $N_{\rm eff}(ab)$ vs.\ $p_{ab}$ (Figs.~\ref{fig:wvspjj} and \ref{fig:wvspSU3}) shows the pair-weight fraction of a configuration, which is global, and the spread of its pairs over channels, which is local. By the chain rule given in Eq.~\eqref{eq:su3HLcd}, the plot resolves the within-configuration term, configuration by configuration, while the mutual information carries the diversity between configurations. Specifically, the weight $p_{ab}$ and the local count $N_{\rm eff}(ab)$ place each configuration in one of four regimes:
\begin{itemize}
\item \emph{Large $p_{ab}$, $N_{\rm eff}(ab)\approx1$:} important but simple; one dominant channel.

\item \emph{Large $p_{ab}$, large $N_{\rm eff}(ab)$:}
important pair weight spread over many channels.

\item \emph{Small $p_{ab}$, large $N_{\rm eff}(ab)$:} locally complex, globally negligible; discarding the configuration removes pair-weight fraction $p_{ab}$,
regardless of the local count.
This is the regime of the weakly populated high-$\eta$ pairs.

\item \emph{Small $p_{ab}$, small $N_{\rm eff}(ab)$:} irrelevant.
\end{itemize}
The boundaries are set by the truncation share $\epsilon$ of Eq.~\eqref{eq:Dtrunc}, that is, ``small'' means below what the discarded weight can absorb. Typically, in the $jj$ scheme most configurations lie on the right, many at large $p_{ab}$ and large $N_{\rm eff}(ab)$ (up to $\approx8$). The reference lies close to the correlated values, except in $^4$He, where it is much higher and extends to the upper left. In SU(3) most configurations sit at large $p_{ab}$ and small $N_{\rm eff}(ab)$, with some at small $p_{ab}$ and intermediate $N_{\rm eff}(ab)$. The reference lies slightly above, again except in $^4$He, where it is visibly higher. SU(3) thus moves the important configurations from the costly
regime into the simple one: per configuration, the pair strength needs fewer channels than in $jj$. The gap to the reference shows that correlation itself concentrates the pairs, most strongly in $^4$He. Ranked by the cumulant $|\Delta p_{ab}|$ instead of $p_{ab}$, the regimes describe the entangling part of the circuit rather than the whole state, and they can differ. In short, $jj$ $N_{\rm eff}$ indicates how many $J$ channels a configuration uses. SU(3) $N_{\rm eff}$ indicates whether the configuration is collective, meaning dominated by its leading irrep, and at which level of the coupling correlation concentrates it.

\section{Channel-guided register layout}
\label{sec:layout}

\begin{figure*}[ht]
    \centering
    \includegraphics[width=0.95\linewidth]{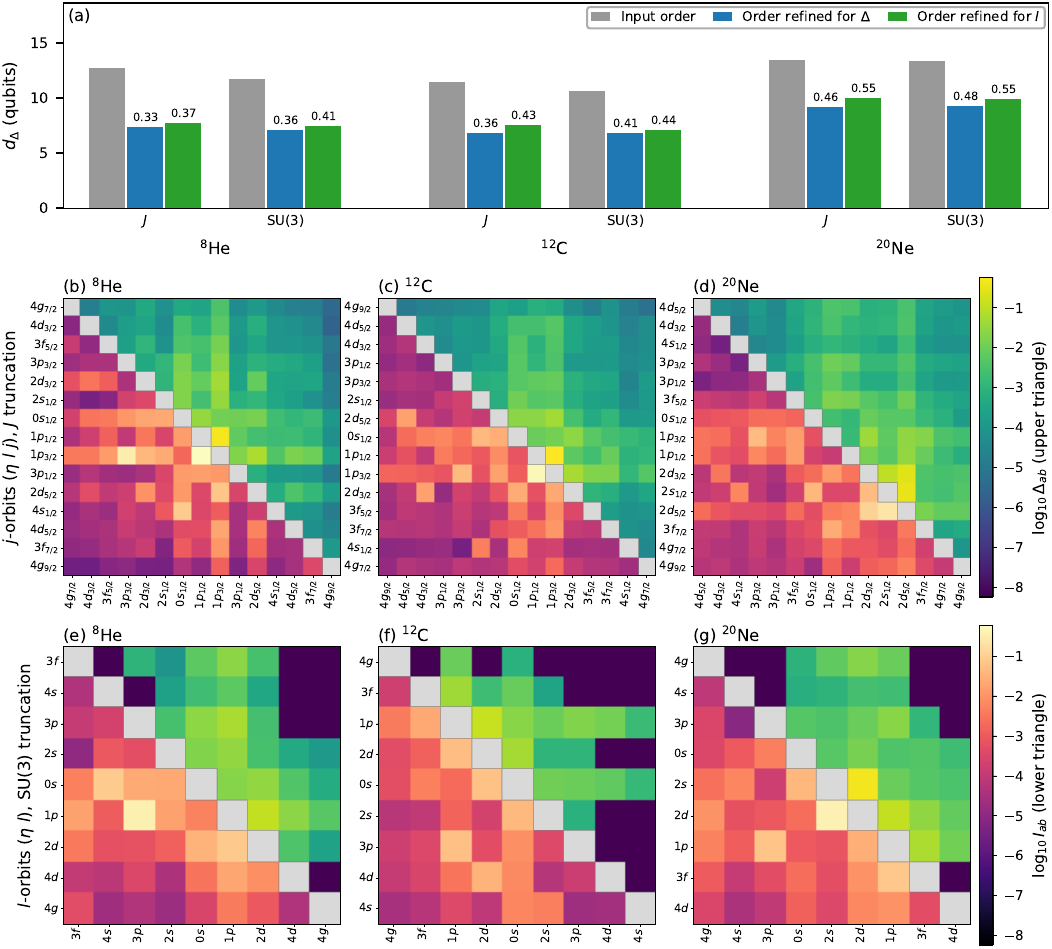}
    \caption{
    Register layout for  $^8$He, $^{12}$C and $^{20}$Ne, using neutron densities and modes with $\eta\le4$. (a) Root-mean-square register distance of the modes linked by the channel graph, $d_\Delta$ (in qubits), for the input order (gray) and the orders refined for $\Delta$ (blue) and for the mutual information $I$ (green), together with the cost relative to the input order, $C_{\rm rel}=(d_\Delta/d_\Delta^{\rm input})^2$ (numbers above the bars). The refined orders are local minima from five starts (input, Fiedler and third-eigenvector orders of $\Delta$ and $I$). $J$ labels $jj$ modes with $j$-orbits as units (in which the order is $m_j=j,-j,(j-1),-(j-1),\dots$) and $\Delta$ truncated to the pair $J$; SU(3): LS modes with $l$-orbits as units (in which the order is $(m_l, m_s)$ followed by $(-m_l, -m_s)$ in decreasing values of $m_l$ and $m_s$) and $\Delta$ truncated to the SU(3) irreps $(\lambda\mu)S$.  (b--g) Unit graphs $\log_{10}\Delta_{ab}$ (upper triangle) and $\log_{10}I_{ab}$ (lower triangle), both of which are summed over the $m$ projections and reported for units, in the order refined for $\Delta$, for (b, e) $^8$He, (c, f) $^{12}$C and (d, g) $^{20}$Ne; values below $10^{-8}$ of the largest, including zeros, are shown in the darkest color.
    }
    \label{fig:registercost}
\end{figure*}

The channels describe which modes are correlated and in what coupled combination. Specifically, in a channel-filtered correlation graph, the weights select the channels first. For a channel $c=(\lambda\,\mu)S\alpha$ or $c=J$, we use Eq.~\eqref{eq:su3dp} and $\Delta p^{c}_{ab} = \sum_M \Delta\Gb^{cM,cM}_{ab,\,ab}$. 
We use the smallest set of channels $(\lambda\,\mu)S$ or $J$ that carries the 99\% fraction of $\sum|\Delta p|$ (Fig.~\ref{fig:Dshare}, solid lines), where all $\alpha$ are kept within the retained  $(\lambda\,\mu)S$ channels.
For two modes $i=(a\,m_a)$ and $j=(b\,m_b)$ in distinct units, $a\neq b$, with
$M=m_a+m_b$, we define:
\begin{equation}
\Delta_{ij}=\Big|\sum_{c,c'}
  C^{cM}_{ij}\,\Delta G^{cM,c'M}_{ab,\,ab}\,C^{c'M}_{ij}\Big|,
\label{eq:Delta}
\end{equation}
where $C^{cM}_{ij}=[R^{cM}_{ab}]_{m_am_b}$ are the coefficients of the pair
state $c$, coupled to $JM$, in the modes $i,j$ [Eq.~\eqref{eq:su3R}]. In the $jj$ scheme, each $J$ has a single pair state, and Eq.~\eqref{eq:Delta} for $J_0=0$ reduces to $\Delta_{ij}=|\sum_{J}(C^{JM}_{ij})^2\Delta G^{JM}_{ab,ab}|$. With this, we can compare the $jj$ and $LS$ coupling schemes for single-particle basis states, since the $LS$ naturally follows the SU(3)-coupling. Specifically, for $jj$ scheme $j$-orbits are used as units and $\Delta$ is truncated to the pair $J$ (Fig.~\ref{fig:Dshare}, blue), while for the $LS$ scheme $l$-orbits are used as units and $\Delta$ is truncated to the SU(3) irreps $(\lambda\mu)S$ (Fig.~\ref{fig:Dshare}, red). 

The excess pair term of two distinct modes relative to the reference, restricted to $c$, defines a weighted graph on the modes of the same species. This is then ordered by its Fiedler vector~\cite{Fiedler1973} and further refined. Compared with ordering on $I_{i,j}$ alone, it shows which channel ties two modes together and omits channels outside the selected candidate pool.

The layout ordering is determined by the Fiedler vector~\cite{Fiedler1973}. Specifically, for a symmetric weight matrix $\Delta_{ij} \ge 0$ (or $I_{i,j}$) the graph Laplacian $L = \Delta' - \Delta$, with $\Delta'$ the diagonal degree matrix $\Delta'_{ii} = \sum_j \Delta_{ij}$, is positive semidefinite with a zero eigenvalue belonging to the constant vector. The eigenvector $\mathbf{f}$ of the second-smallest eigenvalue is the Fiedler vector~\cite{Fiedler1973}, and sorting the modes by its components gives the ordering. It solves:
\begin{equation}
\begin{gathered}
\min_{\mathbf{x}} \sum_{i,j} \Delta_{ij}\,(x_i - x_j)^2\\
\text{subject to}\quad \sum_i x_i = 0,\quad \sum_i x_i^2 = 1,
\end{gathered}
\label{eq:su3fiedler}
\end{equation}
where $x_i$ the position of mode $i$. Eq.~\eqref{eq:su3fiedler} relaxes to continuous positions the 2-sum problem of graph layout, that is, placing strongly correlated modes close together on a line~\cite{JuvanMohar1992,GeorgePothen1997}.  Equation~\eqref{eq:su3fiedler} measures register locality by weighting the squared separation of each pair of modes by its correlation strength. With the mutual information $I_{i,j}$ in place of $\Delta_{ij}$, the same relaxation is the standard way to order orbitals in DMRG calculations~\cite{Legeza2003,Barcza2011}.

For nuclear systems, there are three points that require special care. If the graph splits into components with no weight between them, each is ordered separately. In a $J_0 = 0$ state, modes related by rotation make the Laplacian degenerate and the Fiedler vector non-unique; the remedy is to merge each symmetry unit, a subshell, a $j$-level or a time-reversed pair, into one node, order the units, and fix the internal order by symmetry. And the relaxation is a heuristic: rounding $\mathbf f$ to an order is not guaranteed to lower the discrete cost (the 2-sum problem is NP-complete~\cite{GeorgePothen1997}), and when $\lambda_2\ll\lambda_3$ a weakly attached tail decides $\mathbf f$.

To estimate the goodness of the register layout order and compare the basis coupling scheme, $jj$ and $LS$ (see Fig.~\ref{fig:registercost} for $^8$He, $^{12}$C, and $^{20}$Ne), we introduce the root-mean-square register distance of the modes linked by the channel graph, $d_\Delta=[\sum_{i<j}\Delta_{ij}(x_i-x_j)^2/\sum_{i<j}\Delta_{ij}]^{1/2}$ (in qubits), shown for the input order and the orders refined for $\Delta$ (Fig.~\ref{fig:registercost}, blue) and for the mutual information $I$ (Fig.~\ref{fig:registercost}, green). We also define the cost relative to the input order, $C_{\rm rel}=(d_\Delta/d_\Delta^{\rm input})^2$.

Reordering the neutron register reduces the layout cost by a factor of about $2$--$3$  relative to the input order (the original shell ordering). The reduction increases as more shells are included in the TBDME (for the same wavefunction), with a reduction factor of 10 in the case of  ${}^{4}\mathrm{He}$ for 7 shells. A large factor implies the correlations become localized in a few strongly linked units, whereas a small factor is indicative either of a good input order or uniformly spread correlations. 

The matrices in Fig.~\ref{fig:registercost}(b)--(g) are shown in the order refined for the channel graph $\Delta$. Unlike the mutual information, $\Delta$ is the cumulant of the pair density, from which the Slater-determinant part, including the one-body coherence, has been subtracted, truncated to the pair channels ($J$ or SU(3) irreps) that carry 99\% of it, and so measures the correlations that the entangling part of the circuit must generate. The one-body coherence that dominates $I$ involves no channel truncation and links radial partners. On an oscillator-basis register, it is removed by an orbital rotation to the natural basis. If the state preparation begins with such a rotation, a layer of Givens rotations, the register layout needs to localize only the correlations in $\Delta$. The $\Delta$-refined order places the valence shell at the center of the register, for $^8$He and $^{12}$C the $1p$ orbits ($\eta=1$) together with $0s$, for $^{20}$Ne the $2s$ and $2d$ orbits, with the strongest links within this cluster, and arranges the remaining units around it by link strength, the strongest closest, alternating left and right. The links of this cluster to the other units are about an order of magnitude stronger than the links among those units. With this, the cost of the input order is reduced to $C_{\rm rel}=0.33$--$0.48$ [blue in Fig.~\ref{fig:registercost}(a)]. For comparison, the order refined for $I$ [green in Fig.~\ref{fig:registercost}(a)], which follows parity and the radial chains, is 7--20\% more costly for $\Delta$.

Even though the $J$ and SU(3) orders differ in their units and details, the reduction is similar in both schemes. Both describe the same pair cumulant of the same state, with its strongest links within the valence cluster, so both refinements make the same main change: they move this cluster from its position in the shell order to the center of the register and place the units most strongly linked to it next to it. How the remaining units are arranged is a smaller effect.

\section{Conclusion}
\label{sec:conclusion}

In the present work, we use nuclear symmetries to investigate entanglement and correlations in nuclei using the \textit{ab initio} symmetry-adapted no-core shell model. The analysis combines the symmetry structure of the many-body wavefunctions with mode entanglement and two-nucleon correlations obtained from symmetry-coupled one- and two-body densities. It describes both correlations between single-particle modes and the distribution of like-particle two-nucleon correlations among symmetry channels.

For ${}^{4,6,8}\mathrm{He}$, ${}^{12}\mathrm{C}$, and ${}^{20}\mathrm{Ne}$, single-mode entropies are largest in the valence shells, while two-mode mutual information reveals substantial correlations both within and between shells. The valence shells therefore emerge as a natural starting point for selecting active orbitals, while mutual information identifies their correlations with other shells. In the harmonic-oscillator basis, nonzero negativity identifies entangled mode pairs in ${}^{6}\mathrm{He}$ and ${}^{12}\mathrm{C}$. Natural orbitals generally yield lower mode entropies than harmonic-oscillator orbitals, with smaller differences in the heavier nuclei studied. The natural-orbital occupations provide basis-independent measures of departure from a Slater determinant, distinguishing the underlying correlation from the basis dependence of individual mode entropies. The linear entropy further provides upper and lower bounds on the summed natural-orbital entropy without diagonalizing the one-body density.

At the shell level, filling fractions set upper bounds on shell entropy and inter-shell mutual information. A lower bound on the latter follows from the largest mutual information between modes in the two shells. In particular, a small shell-entropy upper bound implies that the shell is only weakly entangled with the rest of the nucleus, allowing candidates for freezing to be identified from one-body densities alone. In ${}^{20}\mathrm{Ne}$, these bounds identify the nearly filled $0s$ shell and weakly occupied higher shells as such candidates.

The symmetry-channel analysis provides complementary information about the organization of like-particle pair correlations. Comparing configuration weights with their pair-channel entropies and effective channel numbers shows that many strongly weighted configurations concentrate their pair strength in only a few SU(3) channels. Pair occupations describe where this strength resides, whereas cumulants relative to the uncorrelated reference identify changes not explained by the one-body density. Comparing SU(3) and $jj$ coupling shows that the SU(3) organization can retain a prescribed share of the cumulant strength with fewer channels, with the operator-component advantage depending on the nucleus. For proton pairs in ${}^{20}\mathrm{Ne}$, capturing $90\%$ of the cumulant strength requires 8 SU(3) channels, compared with 53 in $jj$ coupling. This example requires fewer channels and operator components, showing a resource advantage for the SU(3)-coupling scheme over $jj$ coupling. The cumulant ranking therefore identifies candidate like-particle pair-excitation operators associated with the strongest correlations beyond the reference.

From the channels selected for the candidate operator set, we construct the correlation graphs that guide quantum register orderings within each species. Local refinement, with Fiedler orderings among the starting points, reduces the correlation-weighted layout cost relative to the input ordering. In the neutron registers studied here, both coupling schemes place strongly linked valence orbits near the center, reducing the $\Delta$-weighted layout cost to $33-48\%$ of that for the original shell ordering. Overall, our results show that nuclear symmetry organizes the dominant like-particle pair correlations into a compact set of channels, extending its computational role from reducing many-body basis dimensions to guiding the choice of excitation operators and quantum register orderings. Applying these selections and register orderings in quantum simulations and assessing their effects on accuracy and computational resources remain subjects for further work. Comparison with SA-NCSM energies and other observables as the selected channel set is enlarged would test whether the dominant correlation channels suffice for an accurate description of the nuclei studied.

\section{Acknowledgments}

We thank Alexis Mercenne for helpful discussions.
This work was supported by the U.S. Department of Energy (DOE), DE-SC0023694. 
This work was performed in part at the Aspen Center for Physics, which is supported by National Science Foundation grant PHY-2210452.
This research benefited from resources of the Oak Ridge Leadership Computing Facility at the Oak Ridge National Laboratory, which is supported by the Office of Science of the U.S. Department of Energy under Contract No. DE-AC05-00OR22725.
The work benefited from high performance computational resources provided by LSU (www.hpc.lsu.edu), the National Energy Research Scientific Computing Center (NERSC), a U.S. Department of Energy Office of Science User Facility at Lawrence Berkeley National Laboratory operated under Contract No. DE-AC02-05CH11231, as well as the Frontera computing project at the Texas Advanced Computing Center, made possible by National Science Foundation award OAC-1818253.
We also thank the Galileo Galilei Institute for Theoretical Physics for the hospitality and the INFN for partial support during the completion of this work.

\section{Data Availability}

The data that support the findings of this study are available from the corresponding author upon reasonable request.

\section{Statement of AI assistance} 

Opus 5.5 assisted with exploration, exposition, literature checks, some figure preparation, manuscript revisions, and numerical tests during the development of this manuscript. The checks performed during this assistance were internal reviews, not independent external peer review or formal verification. The authors defined the research problem and carried forward SU(3)-related mathematical derivations and large-scale calculations. In addition, responsibility for the final manuscript, its mathematical derivations, and its claims rests with the authors.

%
%
\begin{appendix}
\appendix
\begin{widetext}

\section{Unitary Transformation Matrices}

For each pair of subshells $(a,b)$ and channel $(\lambda\,\mu)\kappa(LS)JM$, let $U^{(\lambda\,\mu)\kappa(LS)JM}_{ab}$ be the
$\Oms_a \times \Oms_b$ matrix:
\begin{equation}
\begin{aligned}
\big[U^{(\lambda\,\mu)\kappa(LS)JM}_{ab}\big]_{m_a m_b}
= \sum_{M_LM_S} (-1)^{\eta_b+l_b - m_{l_b} + \frac12 - m_{s_b}}& \cgr{(\eta_a\,0)}{l_a}{\tilde{(0\,\eta_b)}}{l_b}{(\lambda\,\mu)}
{\kappa L}
\cg{l_a}{m_{l_a}}{l_b}{-m_{l_b}}{L}{M_L}\\[-2pt]
\times\;&\cg{\tfrac12}{m_{s_a}}{\tfrac12}{-m_{s_b}}{S}{M_S}\,
\cg{L}{M_L}{S}{M_S}{J}{M}
\end{aligned}
\label{eq:su3U}
\end{equation}
such that,
\begin{equation}
a^\dagger_{\ama}\, a_{\bmb}
= \sum_{(\lambda\,\mu)\kappa LSJM} \big[U^{(\lambda\,\mu)\kappa(LS)JM}_{ab}\big]_{m_a m_b}\; T^{(\lambda\,\mu)\kappa(LS)JM}_{ab},
\label{eq:su3invert}
\end{equation}
and because the three Clebsch-Gordan coefficients implement a unitary recoupling, the
$U$ are orthonormal in the trace inner product:
\begin{equation}
\operatorname{Tr}\big[U^{(\lambda\,\mu)\kappa(LS)JM\dagger}_{ab}\, U^{(\lambda'\,\mu')\kappa'(L'S')J'M'}_{ab}\big]
= \delta_{(\lambda\,\mu)(\lambda'\,\mu')}\delta_{\kappa\kappa'}\delta_{LL'}\,\delta_{SS'}\,\delta_{JJ'}\,\delta_{MM'} .
\label{eq:su3ortho}
\end{equation}
with $[U^{(\lambda\,\mu)\kappa(LS)JM\dagger}_{ab}]_{m_bm_a} = [U^{(\lambda\,\mu)\kappa(LS)JM}_{ab}]^*_{m_am_b}$.
The diagonal case is written as $U^{(\lambda\,\mu)\kappa(LS)JM}_{aa}$, a square $\Oms_a \times \Oms_a$ matrix. Only
$M = 0$ enters expectation values. With $\sum_{m_{l_a}} (-1)^{\,l_a - m_{l_a}}\,
\cg{l_a}{m_{l_a}}{l_a}{-m_{l_a}}{L}{M_L}
\;=\; \Pi_{l_a}\,\delta_{L0}\,\delta_{M_L 0}$, and $\Oms_a=2\Pi_{l_a}^2$ (and similarly for spin):
\begin{equation}
\begin{gathered}
\operatorname{Tr}\,U^{(\lambda\,\mu)\kappa(LS)JM}_{aa} = \sum_{m_a=l_a m_{l_a} m_{s_a}}\big[U^{(\lambda\,\mu)\kappa(LS)JM}_{aa}\big]_{m_a m_a}  \\
=\sum_{l_a} (-1)^{\eta_a} \cgr{(\eta_a\,0)}{l_a}{\tilde{(0\,\eta_a)}}{l_a}{(\lambda\,\mu)}{\kappa L}
\Pi_{l_a}\,\Pi_{1/2}\;\delta_{L0}\,\delta_{S0}\,\delta_{J0}\,\delta_{M0},\\
=\sqrt{\dim_a}\delta_{L0}\,\delta_{S0}\,\delta_{J0}\,\delta_{M0}\sum_{l_a} \frac{(-1)^{\eta_a} \Pi_{l_a}}{\sqrt{\mathrm{dim}(\eta_a,0)}}\cgr{(\eta_a\,0)}{l_a}{\tilde{(0\,\eta_a)}}{l_a}{(\lambda\,\mu)}{\kappa=0 L=0}
\\
=\sqrt{\dim_a}\delta_{(\lambda\,\mu)(0\,0)}\delta_{\kappa 0}\delta_{L0}\,\delta_{S0}\,\delta_{J0}\,\delta_{M0}.
\end{gathered}
\label{eq:su3trU}
\end{equation}
with $U^{(00)0(00)00}_{aa} = \frac{\one}{\sqrt{\dim_a}} $, where we use $\braket{(\eta\,0)l;(0\,\eta)l||(0\,0)\kappa=0 L=0} = (-1)^\eta \Pi_l / \sqrt{\mathrm{dim}(\eta,0)}$ and $$\sum_{l_a l_b} {\cgr{(\lambda_a\,\mu_a)}{l_a}{(\lambda_b\,\mu_b)}{l_b}{(\lambda\,\mu)}{\kappa L}}_{\rho{}}{\cgr{(\lambda_a\,\mu_a)}{l_a}{(\lambda_b\,\mu_b)}{l_b}{(\lambda'\,\mu')}{\kappa L}}_{\rho'}=\delta_{(\lambda\,\mu)(\lambda'\,\mu')}\delta_{\rho,\rho'}.$$ 

\end{widetext}

The expectation value of Eq.~\eqref{eq:su3invert} defines the one-body density matrix for modes:
\begin{equation}
\gamma_{ab} \;=\; \sum_{(\lambda\,\mu)S\alpha} \bar O^{(\lambda\,\mu)S\alpha0}_{ab}\; U^{(\lambda\,\mu)S\alpha0}_{ab}.
\label{eq:su3gamma}
\end{equation}
The full matrix $\gamma = [\gamma_{ab}]$ (with $[\gamma]_{\ama,\,\bmb} = \gamma_{\ama,\,\bmb}$) over all subshells is the one-body particle-reduced density matrix (see L\"owdin~\cite{Lowdin1955} and
Coleman~\cite{Coleman1963}), with trace $A$ and eigenvalues in $[0,1]$. Each diagonal block $\gamma_{aa}$ is a principal submatrix, Hermitian and positive semidefinite with eigenvalues in $[0,1]$ and trace $n_a$, whereas the off-diagonal blocks carry the inter-subshell coherences.

Because the $U$ are orthonormal, Eq.~\eqref{eq:su3gamma} inverts to a trace:
\begin{equation}
\bar O^{(\lambda\,\mu)S\alpha0}_{ab} = \operatorname{Tr}\big[U^{(\lambda\,\mu)S\alpha0\dagger}_{ab}\, \gamma_{ab}\big] 
\label{eq:su3Otrace}
\end{equation}
so each OBDME is the component of the one-body density along one orthonormal
multipole matrix. Combining Eqns.~\eqref{eq:su3gamma} and \eqref{eq:su3Otrace} leads to:
\begin{eqnarray}
\gamma_{\ama}
&=& 
\sum_{\lambda \kappa LSJ} \bar O^{(\lambda\,\lambda)S\alpha0}_{aa}
\big[U^{(\lambda\,\lambda)S\alpha0}_{aa}\big]_{m_a m_a}\cr
&=& 
\sum_{\lambda \kappa LSJ} \operatorname{Tr}\big[U^{(\lambda\,\lambda)S\alpha0\dagger}_{aa}\, \gamma_{aa}\big]\,
\big[U^{(\lambda\,\lambda)S\alpha0}_{aa}\big]_{m_a m_a}.
\label{eq:su3xgeneral}
\end{eqnarray}

Each $U^{(\lambda\,\lambda)(LS)J}_{aa}$ is itself a normalized operator built from the orbital and spin
angular momenta of the subshell. On the $(2l_a+1)$-dimensional orbital space each rank
$L \le 2l_a$ occurs exactly once among the operators, so every rank-$L$ tensor there is a
multiple of one built from $\mathbf{l}$, a polynomial of degree $L$ in its components; on the
spin-$\tfrac12$ space the ranks are $0$ and $1$, carried by $\one$ and $\mathbf{s}$. Hence, we have:
\begin{equation}
\begin{gathered}
U^{(\lambda\,\lambda)S\alpha0}_{aa} \propto T^{(\lambda\,\lambda)S\alpha0}_{aa},\\
T^{(\lambda\,\lambda)S\alpha M}_{aa} = \big\{ t^{(\lambda\,\lambda)L}(\mathbf{l}) \times t^{(0\,0)S}(\mathbf{s}) \big\}^{JM},
\end{gathered}
\label{eq:su3Tops}
\end{equation}
with $t^{(0\,0)0} = \one$, $t^{(1\,1)1}(\mathbf{l}) \propto \mathbf{l}$, $t^{(0\,0)1}(\mathbf{s}) \propto \mathbf{s}$, and
$t^{(\lambda\,\lambda)2}(\mathbf{l}) \propto [\mathbf{l} \times \mathbf{l}]^{(\lambda\,\lambda)2}$, and the most general form of
Eq.~\eqref{eq:su3xgeneral}, valid for any normalization of the operators, is:
\begin{eqnarray}
\;
\gamma_{\ama}
= \sum_{\lambda \kappa LSJ}
\frac{\operatorname{Tr}\big[\gamma_{aa}\, T^{(\lambda\,\lambda)S\alpha0\dagger}_{aa}\big]}
{\operatorname{Tr}\big[T^{(\lambda\,\lambda)S\alpha0\dagger}_{aa}\, T^{(\lambda\,\lambda)S\alpha0}_{aa}\big]}\;
\cr 
\times \big[T^{(\lambda\,\lambda)S\alpha0}_{aa}\big]_{m_a m_a} \;
\label{eq:su3xoperator}
\end{eqnarray}
Each numerator is the one-body expectation value of a subshell SU(3) operator, so a mode occupation is a sum of expectation values of SU(3) operators, weighted by their diagonal matrix elements.

The two-body counterpart of $U$ recouples two creation operators. For each pair of subshells $(a,b)$ and pair channel $(\lambda\,\mu)S\alpha M$, let $R^{(\lambda\,\mu)S\alpha M}_{ab}$ be the
$\Oms_a \times \Oms_b$ matrix:
\begin{equation}
\begin{aligned}
\big[R^{(\lambda\,\mu)S\alpha M}_{ab}\big]_{m_a m_b}
&= 
\cgr{(\eta_a\,0)}{l_a}{{(\eta_b\,0)}}{l_b}{(\lambda\,\mu)}{\kappa L}\cr
&\quad\times 
\cg{l_a}{m_{l_a}}{l_b}{m_{l_b}}{L}{M_L}\cr
&\quad\times
\cg{\tfrac12}{m_{s_a}}{\tfrac12}{m_{s_b}}{S}{M_S}\cr
&\quad\times
\cg{L}{M_L}{S}{M_S}{J}{M}.
\end{aligned}
\label{eq:su3R}
\end{equation}
Like $U$ it is built from Clebsch-Gordan coefficients alone and does not depend on the wavefunction. Inverting the pair operator of Eq.~\eqref{eq:su3pairop},
$P^{(\lambda\,\mu)S\alpha M\dagger}_{ab} = \nu_{ab} \sum_{m_a m_b} [R^{(\lambda\,\mu)S\alpha M}_{ab}]_{m_a m_b}\,
a^\dagger_{\ama} a^\dagger_{\bmb}$ gives:
\begin{equation}
a^\dagger_{\ama}\, a^\dagger_{\bmb}
= \nu_{ab}^{-1} \sum_{LSJM} \big[R^{(\lambda\,\mu)S\alpha M}_{ab}\big]_{m_a m_b}\,
P^{(\lambda\,\mu)S\alpha M\dagger}_{ab},
\label{eq:su3Rinvert}
\end{equation}
with the sum restricted to $L+S$ even for $a = b$. For $a \ne b$ the recoupling is unitary: the $R$ are orthonormal, $\operatorname{Tr}[R^{(\lambda\,\mu)S\alpha M\,T}_{ab} R^{(\lambda'\, \mu')\kappa'(L'S')J'M'}_{ab}] = \delta_{LL'}\delta_{SS'}\delta_{JJ'}\delta_{MM'}$, and complete, $\sum_{LSJM} [R]_{m_a m_b}[R]_{n_a n_b} = \delta_{m_a n_a}\delta_{m_b n_b}$. For $a = b$, exchanging the two particles multiplies $R$ by $(-1)^{L+S+1}$, so the allowed channels, $L+S$ even, are antisymmetric, and their completeness sum is the antisymmetric projector $\tfrac12(\delta_{m_a n_a}\delta_{m'_a n'_a} - \delta_{m_a n'_a}\delta_{m'_a n_a})$. This supplies the factor $\nu_{aa}^{-1} = \sqrt2$ in Eq.~\eqref{eq:su3Rinvert}.

The pair term is quadratic in Eq.~\eqref{eq:su3Rinvert}, so the two-body analogue of the one-body $U$ is a product of two $R$ matrices. It is written $U$ as well, distinguished from the one-body matrix by carrying two channel labels:
\begin{equation}
\begin{aligned}
&\big[U^{(\lambda\,\mu)S\alpha M,\,(\lambda'\, \mu')S'\alpha' M}_{ab}\big]_{m_a m_b}\\
&\quad= (1+\delta_{ab})\,\big[R^{(\lambda\,\mu)S\alpha M}_{ab}\big]_{m_a m_b}
\big[R^{(\lambda'\, \mu')S'\alpha' M}_{ab}\big]_{m_a m_b}
\end{aligned}
\label{eq:su3Upair}
\end{equation}
with $M$ fixed by the two modes. By the completeness relations above, the diagonal weights satisfy $\sum_{(\lambda\,\mu)S\alpha } [U^{(\lambda\,\mu)S\alpha M,\,(\lambda\,\mu)S\alpha M}_{ab}]_{m_a m_b} = 1$ for two distinct modes, both for
$a \ne b$ and, through the factor $(1+\delta_{ab})$, for $a = b$.

\section{Refined order for channel-guided register layout} 

The Fiedler order is therefore a starting point for a local search on the discrete cost:
\begin{equation}
C(\pi) = \sum_{i<j}\Delta_{ij}\,(x_i - x_j)^2 ,
\label{eq:su3cost}
\end{equation}
as in the spectral-sequencing heuristics for linear arrangement~\cite{Petit2003,Diaz2002}, with $x_i$ the register position of mode $i$ when the units (subshells or $j$-levels) are placed in the order $\pi$, each unit contiguous and with its internal order fixed by symmetry. The procedure is as following: (i) The starts are the input (shell) order, the Fiedler orders of $\Delta$ and of $I$, and the orders by the third Laplacian eigenvector of each. (ii) From a start, one unit is removed and reinserted at another position (a move by one position is a swap of neighbors), and the move is accepted if it lowers $C$. (iii) Sweeps over all units and positions are repeated until no move lowers $C$. (iv) The lowest-cost result over the starts is kept. Every accepted move lowers $C$, so the result is never worse than any start, although it is a local, not a global, minimum. With the internal orders fixed, $C$ depends on $\pi$ only through the unit offsets $o_u$, $C = \tfrac12\sum_{u\ne v}\big[S^{(0)}_{uv}d_{uv}^2 + 2S^{(1)}_{uv}d_{uv} + S^{(2)}_{uv}\big] + \text{const}$, with $d_{uv} = o_u - o_v$, $S^{(k)}_{uv} = \sum_{i\in u,\,j\in v}\Delta_{ij}(l_i - l_j)^k$ and $l_i$ the position of mode $i$ inside its unit, so a trial costs $O(n_{\rm units}^2)$ operations. The same search is run with $I_{i,j}$ in place of $\Delta_{ij}$.

For SU(3), $S^{(0)}_{uv}$, the total weight between two units, is the leading term and the only one that needs no mode resolution; $S^{(1)}_{uv}$ says whether the correlated modes of $u$ sit at the end that faces $v$, and is where the magnetic labels enter; $S^{(2)}_{uv}$ does not depend on the order. At the shell level, $S^{(0)}\to\Delta_{\eta_a\eta_b}$ and the weight is spread uniformly over the mode pairs of the two blocks, which sets $S^{(1)}=0$:
\begin{equation}
C(\pi)=\sum_{\eta_a<\eta_b}\Delta_{\eta_a\eta_b}\Big[(c_{\eta_a}-c_{\eta_b})^2+\frac{\Om_{\eta_a}^2-1}{12}+\frac{\Om_{\eta_b}^2-1}{12}\Big],
\label{eq:shellcost}
\end{equation}
with $c_\eta$ the centre of the block of shell $\eta$ in the order $\pi$. With up to nine shells every order is tried, so the minimum is exact; above that the local search of \S\ref{sec:layout} is used, from the input, Fiedler and third-eigenvector orders.

\section{One-body and two-body reduced densities}

For completeness, we show the one-body and two-body reduced densities, following Ref.~\cite{Robin2021}. The single-orbital reduced density matrix is given by:
\begin{eqnarray}
    \rho_{n_{i},n_{i}'}^{(i)}
    &=&
    \sum_{BC}\bra{n_{i}}\braket{BC|\varPsi} \braket{\varPsi|BC} \ket{n_{i}'}\cr
    &=&
    \sum_{BC}\braket{\varPsi|BC} \ket{n_{i}'}\bra{n_{i}}\braket{BC|\varPsi},
\label{eq:SBS_OBDM}
\end{eqnarray}
where $BC\equiv(n_{1}n_{2}...n_{i-1}n_{i+1}...n_{N})$. Each fixed state $i$ can be occupied or empty so that we have a basis $\left\{ \ket{n_{i}}\right\} =\left\{ \ket{0};\ket{1}= a_{i}^{\dagger} \ket{0}\right\}$. In this basis there are four matrix elements:
\begin{eqnarray}
    \rho_{1,1}^{(i)} 
    &=&
    \sum_{BC}\braket{\varPsi|BC} \ket{1}\bra{1}\braket{BC|\varPsi},\cr
    &=&
    \sum_{BC}\braket{\varPsi|BC} a_{i}^{\dagger}\ket{0}\bra{0}a_{i}\braket{BC|\varPsi},\cr
    &=&
    \sum_{BC}\braket{\varPsi|BC} a_{i}^{\dagger}\left[\sum_{n_{i}}\ket{n_{i}}\bra{n_{i}}-\ket{1}\bra{1}\right]a_{i}\braket{BC|\varPsi},\cr
    &=&
    \sum_{BC}\braket{\varPsi|BC} a_{i}^{\dagger}a_{i}\braket{BC|\varPsi},\cr
    &=&
    \bra{\varPsi}a_{i}^{\dagger}a_{i}\ket{\varPsi}=\gamma_{ii}.\\\cr\cr
\label{eq:OBD_ME1}
    \rho_{0,0}^{(i)} 
    &=&
    \sum_{BC}\braket{\varPsi|BC} \ket{0}\bra{0}\braket{BC|\varPsi},\cr
    &=&
    \sum_{BC}\braket{\varPsi|BC} \left[\sum_{n_{i}}\ket{n_{i}}\bra{n_{i}}-\ket{1}\bra{1}\right]\braket{BC|\varPsi},\cr
    &=&
    1-\gamma_{ii},
\label{eq:OBD_ME2}
\end{eqnarray}
and $\rho_{1,0}^{(i)}=\rho_{0,1}^{(i)}=0$ due to conservation
of particle number. The matrix in this basis can then be written:
\begin{equation}
    \rho^{(i)}=\left(\begin{array}{cc}
    1-\gamma_{ii} & 0\\
    0 & \gamma_{ii}
    \end{array}\right).
\label{OBD_matrix}
\end{equation}

The two-orbital reduced density matrix is given by:
\begin{eqnarray}
    \rho_{n_{i},n_{j},n_{i}',n_{j}'}^{(ij)}
    &=&
    \sum_{C}\bra{n_{j}n_{i}} \braket{C|\varPsi} \braket{\varPsi|C} \ket{n_{i}'n_{j}'}\cr
    &=&
    \sum_{BC} \braket{\varPsi|C} \ket{n_{i}'n_{j}'}\bra{n_{j}n_{i}} \braket{C|\varPsi},
\label{SBS_TBDM}
\end{eqnarray}
where C$\equiv(n_{1}n_{2}...n_{i-1}n_{i+1}...n_{j-1}n_{j+1}...n_{N})$. There are four states for the basis $\ket{n_{i}n_{j}}$ that we denote:
\begin{equation}
    \ket{\underline{1}}\equiv\ket{00},\,\,\,\,\,\,\ket{\underline{2}}\equiv\ket{01},\,\,\,\,\,\,\ket{\underline{3}}\equiv\ket{10},\,\,\,\,\,\,\ket{\underline{4}}\equiv\ket{11}.
\label{eq:TBDM_states}
\end{equation}
The matrix elements of $\rho^{(ij)}$ are then:
\begin{eqnarray}
    \rho_{4,4}^{(i)} 
    &=&
    \sum_{C} \braket{\varPsi|C} \ket{11}\bra{11} \braket{C|\varPsi},\cr
    &=&
    \sum_{C} \braket{\varPsi|C} a_{i}^{\dagger}a_{j}^{\dagger}\ket{00}\bra{00}a_{j}a_{i} \braket{C|\varPsi},\cr
    &=&
    \sum_{C} \braket{\varPsi|C} a_{i}^{\dagger}a_{j}^{\dagger}\cr
    &\quad&\times
    \left[\sum_{n_{i}}\ket{n_{i}n_{j}}\bra{n_{j}n_{i}}-\ket{01}\bra{10}-\ket{10}\bra{01}-\ket{11}\bra{11}\right]\cr
    &\quad&\times 
    a_{j}a_{i} \braket{C|\varPsi},\cr
    &=&
    \sum_{C} \braket{\varPsi|C} a_{i}^{\dagger}a_{j}^{\dagger}a_{j}a_{i} \braket{C|\varPsi},\cr
    &=&
    \bra{\varPsi}a_{i}^{\dagger}a_{j}^{\dagger}a_{j}a_{i}\ket{\varPsi}=\gamma_{ijij}.\\\cr\cr
\label{eq:TBD_ME1}
    \rho_{3,3}^{(i)}
    &=&
    \sum_{C} \braket{\varPsi|C} \ket{10}\bra{01} \braket{C|\varPsi},\cr
    &=&
    \sum_{C} \braket{\varPsi|C} a_{i}^{\dagger}\ket{00}\bra{00}a_{i} \braket{C|\varPsi},\cr
    &=&
    \sum_{C} \braket{\varPsi|C} a_{i}^{\dagger}\cr
    &\quad&\times
    \left[\sum_{n_{i}}\ket{n_{i}n_{j}}\bra{n_{j}n_{i}}-\ket{01}\bra{10}-\ket{10}\bra{01}-\ket{11}\bra{11}\right]\cr
    &\quad&\times
    a_{i} \braket{C|\varPsi},\cr
    &=&
    \bra{\varPsi}a_{i}^{\dagger}a_{i}\ket{\varPsi}-\sum_{C} \braket{\varPsi|C} \ket{11}\bra{11} \braket{C|\varPsi},\cr
    &=&
    \gamma_{ii}-\gamma_{ijij}.
\label{eq:TBD_ME2}
\end{eqnarray}
Similarly, $\rho_{2,2}^{(i)}=\gamma_{jj}-\gamma_{ijij}$ and $\rho_{1,1}^{(i)}=1-\gamma_{ii}-\gamma_{jj}+\gamma_{ijij}$. Finally, the nonzero off-diagonal elements are $\rho_{2,3}^{(i)}=\gamma_{ji}$ and $\rho_{3,2}^{(i)}=\gamma_{ij}$. All other matrix elements cancel due to particle-number conservation. Hence, the matrix in this basis can be written as in Eq. (\ref{eq:su3Gammadef}).

\end{appendix}
\bibliography{entanglement_refs}
\end{document}